\documentclass[a4paper,12pt]{article}
\usepackage{jheppub,esint,psfrag}
\usepackage[utf8]{inputenc}

\usepackage{epsfig,amssymb,amsmath,psfrag,subfigure,rotate,color,wasysym,xcolor}
\usepackage[bbgreekl]{mathbbol}

\allowdisplaybreaks 
\newcommand{\insertfig}[2]{\includegraphics[width=#1cm]{#2}}

\DeclareSymbolFontAlphabet{\mathbbm}{bbold}
\DeclareSymbolFontAlphabet{\mathbb}{AMSb}%

\def\XXint#1#2#3{{\setbox0=\hbox{$#1{#2#3}{\int}$ }
\vcenter{\hbox{$#2#3$ }}\kern-.6\wd0}}

\def \be  {\begin{equation}}
\def \ee  {\end{equation}}
\def \ba  {\begin{eqnarray}}
\def \ea  {\end{eqnarray}}
\def \baa {\begin{eqnarray*}}
\def \eaa {\end{eqnarray*}}
\newcommand{\ep}{\varepsilon}
\def \lab #1 {\label{#1}}

\newcommand\re[1]{(\ref{#1})}
\def\d{\hbox{{d}\kern-.20em\hbox{l}}}

\def \matrix #1 {\left(\begin{array}{cc} #1 \end{array}\right)}

\def \tr {\mathop{\rm tr}\nolimits}

\newcommand \widebar [1] {\overline{#1}}

\newcommand \vev [1] {\langle{#1}\rangle}
\newcommand \VEV [1] {\left\langle{#1}\right\rangle}
\newcommand \ket [1] {|{#1}\rangle}
\newcommand \bra [1] {\langle {#1}|}

\def\1{\hbox{{1}\kern-.25em\hbox{l}}}

\newcommand{\ft}[2]{{\textstyle\frac{#1}{#2}}}
\makeatletter

\input pdf-trans
\newbox\qbox
\def\usecolor#1{\csname\string\color@#1\endcsname\space}
\newcommand\bordercolor[1]{\colsplit{1}{#1}}
\newcommand\fillcolor[1]{\colsplit{0}{#1}}
\newcommand\outline[1]{\leavevmode%
  \def\maltext{#1}%
  \setbox\qbox=\hbox{\maltext}%
  \boxgs{Q q 2 Tr \thickness\space w \fillcol\space \bordercol\space}{}%
  \copy\qbox%
}
\makeatother
\newcommand\colsplit[2]{\colorlet{tmpcolor}{#2}\edef\tmp{\usecolor{tmpcolor}}%
  \def\tmpB{}\expandafter\colsplithelp\tmp\relax%
  \ifnum0=#1\relax\edef\fillcol{\tmpB}\else\edef\bordercol{\tmpC}\fi}
\def\colsplithelp#1#2 #3\relax{%
  \edef\tmpB{\tmpB#1#2 }%
  \ifnum `#1>`9\relax\def\tmpC{#3}\else\colsplithelp#3\relax\fi
}
\bordercolor{black}
\fillcolor{white}
\def\thickness{.3}

\def\1{\mathbbm{1}}

\title{Towards six W-boson amplitude at two loops}

\author[a]{A.V.~Belitsky}
\affiliation[a] {Department of Physics, Arizona State University,  Tempe, AZ 85287-1504, USA}  
    
 \abstract
{We construct the planar integrand of the six-leg amplitude of massive W-bosons on the special Coulomb branch of the maximally 
supersymmetric Yang-Mills theory to two-loop order. We use the six-dimensional supersymmetric spinor-helicity formalism and the 
generalized unitarity-cut sewing technique to perform this analysis. The thus-found expression corresponds in the massless 
limit to the six-gluon maximally helicity-violating amplitude. The upshot of the current consideration is that the former is just an uplift 
of the latter from four to six dimensions. This repeats the pattern found earlier for four- and five-leg amplitudes of massive W-bosons.}

\begin{document}

\maketitle
\flushbottom
\setcounter{footnote} 0

\section{Introduction}

Scattering amplitudes bridge the gap between experimental data and the theoretical description of Nature. However, their calculation 
from gauge-field Lagrangians is an insurmountable task beyond one-loop order and multiplicity of final-state particles higher than two. 
Just the very generation and summation of a factorially growing number of Feynman diagrams is nearly hopeless, even without factoring 
in the necessity of calculating emerging divergent loop momentum integrals formed by virtual states. The problem becomes even more 
daunting if the external states are off their mass shell. How does one preserve gauge invariance of corresponding amplitudes without 
aggravating the just-mentioned first step? 

The mid-1990s witnessed a practical solution to the predicament mentioned above in the form of a generalized unitarity-cut construction 
of integral bases for loops \cite{Bern:1994zx,Bern:1994cg,Bern:2004cz}. According to it, one builds integrands by sewing together tree 
amplitudes to form amplitude's discontinuities and then merging them together. This accounts for a portion of the result, which reflects the 
branch-cut analytic structure of loops. Possible additive contributions from rational functions of external kinematical invariants can be 
reconstructed as well \cite{Berger:2006cz}. Gauge invariance of on-shell trees ensures the same property for loops. The proficiency of the 
method can be further enhanced if the trees are implemented in terms of a spinor-helicity formalism. The latter is eagerly available 
in four space-time dimensions \cite{Berends:1981rb,DeCausmaecker:1981wzb,Kleiss:1985yh,Gunion:1985vca,Xu:1986xb}. However, the 
restriction to exactly four dimensions is a nuisance since, in massless theories, loop effects develop infrared singularities when virtual 
momenta are soft or collinear to external momenta and require a regulator\footnote{Of course, there are ultraviolet divergences as well
in models describing Nature.}. A form of dimensional regularization is employed as a remedy in gauge theories: it preserves the underlying 
gauge invariance by analytically continuing all expressions to an $\varepsilon$-vicinity of four dimensions, $D = 4 - 2 \varepsilon$. However, 
it forces one to abandon the advantages of the four-dimensional spinor-helicity language, thus setting one a step back. There is, nevertheless, 
a way out of this. Four dimensions do not own spinor-helicity. The latter is available in other space-times, particularly, in $D = 6$ 
\cite{Cheung:2009dc,Dennen:2009vk}. Then, one performs calculations in integer dimensions and constructs an interpolation between 
them \cite{Giele:2008ve} or properly adjusts out-of-four-dimensional degrees of freedom \cite{Bern:2010qa}. This allows one to recover 
loop amplitudes fully, including the so-called $\mu$-terms \cite{Bern:2002tk}.

In this work, we adhere to the extra-dimensional point of view for a different purpose: the study of a gauge-invariant framework to off-shell 
scattering. This point of view was adopted before to introduce mass effects by embedding them in extra-dimensional components of 
massless six-dimensional momenta in a Yang-Mills theory \cite{Selivanov:1999ie,Bern:2002tk,Boels:2010mj,Craig:2011ws}. This vantage 
point is particularly fruitful in $\mathcal{N} = 4$ super-Yang-Mills (sYM), which arises from the dimensional reduction of ten-dimensional 
$\mathcal{N} = 1$ sYM \cite{Brink:1976bc,Gliozzi:1976qd}. It is equivalent to the purely four-dimensional mass generation via a Higgs 
mechanism \cite{Alday:2009zm}. The latter moves the model away from the origin of its moduli space onto its so-called Coulomb branch. 
Our current interest is in a special configuration of vacuum expectation values, ergo extra-dimensional momenta, where only external 
particles are endowed with masses, which we interpret as the off-shellness, while all states propagating in loops are left massless. 

Amplitudes on the special Coulomb branch display infrared properties far removed from their on-shell counterparts. For generic values
of the off-shellness, they are indeed finite. However, as one dials down its magnitude, the logarithms of external states' virtualities plague 
loop effects. While this very fact is no different from massless infrared divergences, their actual all-order structure is quite distinct. This 
was demonstrated for four- \cite{Caron-Huot:2021usw} and five-leg \cite{Bork:2022vat,Belitsky:2025bgb} amplitudes for scattering of 
W-bosons, as well as their various form factors \cite{Belitsky:2022itf,Belitsky:2023ssv,Belitsky:2024agy,Belitsky:2024dcf}. The most 
notable deviation of the Coulomb `observables' as compared to their conformal counterpart is that the Sudakov logarithms are governed by 
the octagon \cite{Belitsky:2019fan} rather than the usual cusp anomalous dimension \cite{Polyakov:1980ca,Korchemsky:1987wg}.

In this work, we construct the integrand of the amplitude for six W-bosons to two-loop order, specifically, the one which reduces to
the maximally helicity-violating configuration when it is taken to its mass shell. The amplitude for six W-bosons will be cast in the form
\begin{align}
\label{LoopAmplitudeGeneric}
{\rm Amplitude}^{(\ell)} = {\rm Amplitude}^{(0)}  \sum_{\alpha \in {\rm Basis}} {\rm coeff}^{(\ell)}_\alpha {\rm Integral}^{(\ell)}_{\alpha}
\, ,
\end{align}
at $\ell$-th perturbative order. At the origin of the moduli space, the basis of integrals defining it was constructed in the seminal paper 
\cite{Bern:2008ap}, making use of the iterated two-particle cuts\footnote{It is worth pointing out that this consideration was immediately 
generalized to any number of legs in \cite{Vergu:2009zm,Vergu:2009tu}.}. It was reconfirmed in the calculations of its leading singularities 
in Ref.\ \cite{Cachazo:2008hp}, the BCFW-like recursion relations \cite{Arkani-Hamed:2010zjl}, as well as heltacuts in \cite{Larsen:2012sx}.

This infrared-sensitive `observable' is of great interest since it is the first multiplicity where one encounters the so-called nontrivial 
{\sl remainder function} \cite{Alday:2007he}, \cite{Bern:2008ap,Drummond:2008aq,Goncharov:2010jf}. The Coulomb branch of 
$\mathcal{N} = 4$ sYM possesses hidden, in addition to the explicit Noether, symmetries. These act in a `dual' space spanned by 
the regions' momenta \cite{Alday:2007hr,Drummond:2006rz,Drummond:2007aua}. While at the conformal point, they are broken by 
divergences intrinsic to quantum effects \cite{Drummond:2007au}; they are preserved on the special Coulomb branch, even when one 
approaches infinitesimally close to the origin of the moduli space. The tree amplitudes, ${\rm Amplitude}^{(0)}$, were shown to be 
covariant under the dual conformal symmetry transformations \cite{Dennen:2010dh,Huang:2011um,Plefka:2014fta}, and this property 
is inherited by quantum loops as well. When the former is pulled out from loops as an overall factor, as in Eq.\ \re{LoopAmplitudeGeneric}, 
the loop integrals ${\rm Integral}^{(\ell)}_{\alpha}$ (with accompanying coefficients) are invariant under dual inversions. This implies that 
at the function level, they depend solely on conformal cross-ratios. Below six external legs, loop amplitudes are functions of the so-called 
`small' conformal cross-ratios, i.e., they vanish uniformly with decreasing external off-shelleness. For six legs, these depend in addition 
on `large' ones, which take generic values even if one considers strictly on-shell kinematics. The remainder function, as alluded to 
above, depends only on these.  At the conformal point, it was of high theoretical value since it provided valuable information about 
the higher-order structure of perturbative series, evidence \cite{Bern:2008ap,Cachazo:2008hp} for the deviation from the 
Bern-Dixon-Smirnov ansatz \cite{Bern:2005iz}, and agreement \cite{Drummond:2008aq} with the dual null polygonal Wilson loop 
\cite{Alday:2007he,Drummond:2007cf,Brandhuber:2007yx}. It even offered an inspiration for a nonperturbative framework via its 
near-collinear expansions \cite{Basso:2013vsa}. 

Presently, we seek this information for the Coulomb amplitude. At four \cite{Caron-Huot:2021usw} and five \cite{Belitsky:2025bgb} 
legs, they are related to correlation functions of infinitely-heavy half-BPS operators \cite{Coronado:2018ypq,Coronado:2018cxj}. 
The question is whether this duality persists for six points. While we will not be able to answer it just yet, presently we make the 
first step towards it. For the special Coulomb branch, we will use a hybrid approach, i.e., employing both the minimal, i.e., iterated-cut, 
unitarity \cite{Bern:2008ap}, and the method of leading singularities \cite{Cachazo:2008hp}. The choice is driven by the efficiency and 
prudence in extracting the analytical form of the expansion coefficients ${\rm coeff}^{(\ell)}_\alpha$. The calculation will be performed 
within $\mathcal{N} = (1,1)$ sYM in six dimensions \cite{Dennen:2009vk}. This theory is not chiral. So, with the chosen extraction 
procedure, which preserves six-dimensional covariance, we will be only sensitive to the sum of chiral and non-chiral contributions. 
Thus, we will not be able to determine parity-odd effects. So our focus is on the parity-even portion of the amplitude only.

Our presentation is organized as follows. In the next section, we recall the form of tree superamplitudes in $\mathcal{N} = (1,1)$ sYM. Then in
Sect.\ \ref{SingletAmplSection}, we focus on the so-called `singlet' component of the six-leg amplitude, which is particularly well suited for
use in the unitarity-cut construction. After that, we move on to planar loops in Sect.\ \ref{LoopSection}. First, in Sect.\ \ref{1LoopSection}, we 
determine the integrand of the one-loop MHV-like six-dimensional amplitude, and then at two loops in Sect.\ \ref{2LoopSection}. Our result 
for a basis of integrals and their accompanying coefficients \re{LoopAmplitudeGeneric} is quoted in Sect.\ \ref{IntBasisSection}. Finally, we 
conclude. A few appendices summarize the basics of the six-dimensional Dirac algebra, as well as the calculation of iterated unitarity cuts 
and leading-singularity analysis at the conformal point in four dimensions.

\section{Six-dimensional trees}

The six-dimensional $\mathcal{N} = (1,1)$ sYM is a non-chiral theory \cite{Dennen:2009vk,Huang:2011um}. All of its on-shell states 
are accessible through the chiral $Q$ and chiral-conjugate $\bar{Q} = {\rm cc} (Q)$ charges. They are classified according to the 
SU$(2) \times$SU$(2)$ little group, which leaves invariant the bosonic momentum $P$ of a particle traveling along the fifth axis. The 
notion of helicity does not exist in the model. It is a curse in disguise. The obvious disadvantage is that if one insists on six-dimensional 
covariance in all calculations, one loses a transparent relation to four-dimensional physics. The advantage, however, is that all 
color-ordered amplitudes of a given multiplicity are encoded in a single superamplitude. There is no break-up according to the total 
helicity of all particles involved in scattering, i.e., MHV, NMHV, etc. Thus, expressions are quite compact, provided one uses a proper 
language to write them. Imposing super-momentum conservation, we can extract it in terms of the bosonic\footnote{Below, we, however, 
strip the bosonic Dirac delta along with the accompanying factor $i (2 \pi)^6$ from all expressions. Momentum conservation will 
nevertheless be tacitly implied.} and fermionic delta functions
\begin{align}
\mathcal{A}_n 
\label{GenericAn}
=
i (2 \pi)^6 
\delta^{(6)} \left( P_{12 \dots n} \right) 
\delta^{(4)} \left( Q_{12 \dots n} \right)
\delta^{(4)} \left( \bar{Q}_{12 \dots n} \right)
\widehat{\mathcal{A}}_n 
\, .
\end{align} 
Here and everywhere below, whenever we encounter single-particle characteristics like momentum and supercharges with multiple 
labels, we imply summation over these, e.g., 
\begin{align}
\label{SumConvention}
P_{12 \dots n} \equiv P_1 + P_2 + \dots + P_n
\, , \qquad
Q_{12 \dots n} \equiv Q_1 + Q_2 + \dots + Q_n
\, .
\end{align}
The reduced color-ordered amplitudes $\widehat{\mathcal{A}}_n$ are homogeneous polynomials of degree $n - 4$ in the Grassmann 
variables $Q$ and $\bar{Q}$. The non-chiral nature of the theory implies that $\widehat{\mathcal{A}}_n$ can be chosen to be polynomials 
of equal degrees $[n/2] - 2$ both in the chiral and anti-chiral degrees of freedom \cite{Plefka:2014fta}. The four- and five-leg reduced 
amplitudes are particularly compact \cite{Cheung:2009dc,Dennen:2009vk,Plefka:2014fta}, while the six-leg one is far more involved 
\cite{Belitsky:2025pnw},
\begin{align}
\label{A4}
\widehat{\mathcal{A}}_4^{(0)} 
&
= \frac{1}{S_{12} S_{23}}
\, , \\
\label{CompactA5}
\widehat{\mathcal{A}}_5^{(0)} 
&
=
\frac{- \Omega_{12345}}{S_{12} S_{23} S_{34} S_{45} S_{51}}
\, , \\
\label{CompactA6}
\widehat{\mathcal{A}}_6^{(0)}
&
=
\frac{\mathcal{N}_6}{S_{12} S_{23} S_{34} S_{45} S_{56} S_{61} S_{612} S_{123} S_{234} (U_1 + U_2 + U_3 - 3)} 
\, .
\end{align} 
The four- and five-leg amplitudes possess only two-particle poles in Mandelstam-like invariants $S_{ij} = (P_i + P_j)^2$, while the one 
with six leg develops three-particle $S_{ijk} = (P_i + P_j + P_k)^2$ factorization channels as well. Upon dimensional reduction to four 
space-time dimensions, the latter will only survive in its NMHV helicity configurations. $\widehat{\mathcal{A}}_6^{(0)}$ develops a 
spurious pole at $U_1 + U_2 + U_3 = 3$ in the conformal cross ratios
\begin{align}
\label{CrossRatios}
U_1 = \frac{S_{12} S_{45}}{S_{123} S_{612}}
\, , \qquad
U_2 = \frac{S_{23} S_{56}}{S_{234} S_{123}}
\, , \qquad
U_3 = \frac{S_{34} S_{61}}{S_{612} S_{234}}
\, .
\end{align}
It is an unfortunate nuisance. It disappears only after specific components are extracted from the above generating function upon 
expansion in the Grassmann variables. The numerator in Eq.\ \re{CompactA6} was constructed in Ref.\ \cite{Belitsky:2025pnw} to 
be
\begin{align}
\label{N6numeratorFinal}
\mathcal{N}_6
&
=
\frac{S_{12} S_{61}}{S_{612}} \Omega_2 \Omega_3 - 2 S_{12} \Omega_2 \Omega_4 + \ft12 S_{123} (U_1 + U_2) \Omega_2 \Omega_5
+
\mbox{cyclic permutations}
\, .
\end{align}
Its most concise representation was found in the language of the region's supermomenta $(X, {\mit\Theta}, \bar{\mit\Theta})$,
\begin{align}
P_i = X_{i,i+1}
\, , \qquad
Q_i = {\mit\Theta}_{i,i+1}
\, , \qquad
\bar{Q}_i
=
\bar{\mit\Theta}_{i,i+1}
\, .
\end{align} 
Here, for the dual variables in the right-hand sides, we use a different label convention, $X_{ij} \equiv X_i - X_j$ and ${\mit\Theta}_{ij} \equiv 
{\mit\Theta}_i - {\mit\Theta}_j$, as opposed to Eq.\ \re{SumConvention}. $\mathcal{N}_6$ was built with the help of the very same 
$\Omega$-structures that first made their appearance in the five-leg case \cite{Plefka:2014fta},
\begin{align}
\label{SixElements}
\{ \Omega_1, \Omega_2, \Omega_3, \Omega_4, \Omega_5, \Omega_6 \}
\equiv
\{ \Omega_{12345}, \Omega_{23456}, \Omega_{34561}, \Omega_{45612}, \Omega_{56123}, \Omega_{61234} \}
\, .
\end{align}
The individual $\Omega$'s are defined by the ${\rm SU}^\ast(4)$-invariant inner product 
\begin{align}
\label{MinimalElements}
\Omega_{ijklm} = \ft12  \bra{B_{i,jl}} \bar{B}_{iKm}] + {\rm cc}
\, ,
\end{align}
of dual-conformal covariant bras and kets\footnote{The cc-operation is defined by changing all unbarred symbols to barred ones and 
vice versa.},
\begin{align}
\label{BconfCovs}
\bra{B_{i,jk}} \equiv \bra{{\mit\Theta}_{ij}} \bar{X}_{jk} X_{ki} + \bra{{\mit\Theta}_{ik}} \bar{X}_{kj} X_{ji}
\, , \qquad
|\bar{B}_{i,jk}] \equiv - {\rm cc} \big( \bra{B_{i,jk}} \big)
\, .
\end{align}
The dependence of the accompanying coefficients on Mandelstam-like invariants was then fixed from a highly restrictive collinear 
bootstrap \cite{Belitsky:2025pnw}. 

In dual variables, all tree amplitudes, in particular \re{A4}--\re{CompactA6}, transform covariantly under the dual inversion
\cite{Bern:2010qa,Dennen:2010dh,Plefka:2014fta}
\begin{align}
\label{ConformalInversionAhat}
\mathcal{I} \widehat{\mathcal A}^{(0)}_n = X_1^2 \dots X_n^2 \widehat{\mathcal A}^{(0)}_n
\, .
\end{align}
with the same weight for each leg. This property will be preserved at the integrand level through the unitary-cut construction of loops.

\begin{figure}[t]
\begin{center}
\mbox{
\begin{picture}(0,75)(160,0)
\put(0,0){\insertfig{11}{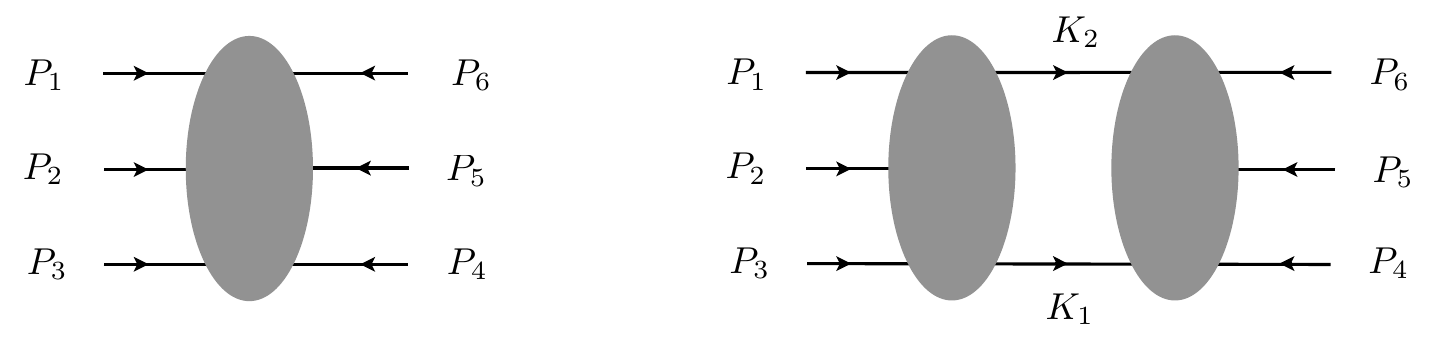}}
\end{picture}
}
\end{center}
\caption{\label{TreeOneLoopPic} Six-leg tree amplitude (left panel) and one-loop unitarity cut (right panel).}
\end{figure}

Eq.\ \re{CompactA6} is the starting point of our subsequent consideration. Throughout the work, we use the `all-in' conventions by 
assuming that all lines in the amplitude are incoming, as demonstrated in the left panel of Fig.\ \ref{TreeOneLoopPic}.

\section{`Singlet' amplitude}
\label{SingletAmplSection}

%%%%%%%%%%%%%%%%%%%%%%%%%%%%%%%%%%%%%%%%%%%%%%%%%%%%%%%%%%%%%%%%%%%%%
%            Figure
%%%%%%%%%%%%%%%%%%%%%%%%%%%%%%%%%%%%%%%%%%%%%%%%%%%%%%%%%%%%%%%%%%%%%
%%%%%%%%%%%%%%%%%%%%%%%%%%%%%%%%%%%%%%%%%%%%%%%%%%%%%%%%%%%%%%%%%%%%%

The complicated Grassmannian structure and, what is worse, the presence of the spurious pole in the six-leg amplitude \re{CompactA6} 
make it very difficult to work with practically. Therefore, we will use a specific projection that will allow us to extract loop integrands most 
compactly and efficiently. We will be able to accomplish this for a component that we dubbed `{\sl singlet}' in Ref.\ \cite{Belitsky:2024rwv}. 
The latter is distinguished by the fact that only one {\sl scalar} configuration will propagate in the intermediate states of a given cut. This is 
akin to {\sl gluon} intermediate states employed in Refs.\ \cite{Bern:2007ct,Bern:2008ap}.

To perform these projections in a covariant fashion, we need to introduce relevant spinor-helicity variables \cite{Cheung:2009dc,Dennen:2009vk}.
Namely, an SU$^\ast(4)$ sextet of on-shell massless particle momenta can be written
\begin{align}
\label{SHmomentum}
P_i = \ket{i^{a}} \bra{i_a}
\, , \qquad
\bar{P}_i = |i_{\dot{a}}][i^{\dot{a}}|
\, ,
\end{align}
in terms of two sets of Weyl spinors $\bra{i^a}$ and $[i_{\dot{a}}|$, which are enumerated by the little group indices for each of the two factors of 
the SU(2) group. Similarly, the (anti-)chiral charges are decomposed via the fermionic variables $\eta_{i,a}/\bar{\eta}_i^{\dot{a}}$ as
\begin{align}
Q_i = \bra{i^a} \eta_{i,a}
\, , \qquad
\bar{Q}_i = [i_{\dot{a}}| \bar{\eta}_i^{\dot{a}}
\, .
\end{align}

Then, for the five-leg amplitude, the `singlet' component is extracted by imposing the following conditions on its chiral charges
\begin{align}
\label{5LegSingletConds}
{\rm sing}
=
\left\{
Q_4 = Q_5 = 0
\, , 
Q_1 = - \frac{Q_{3} \bar{P}_2 P_1}{S_{12}}
\, , 
Q_2 = - \frac{Q_{3} \bar{P}_1 P_2}{S_{12}}
\right\}
\, ,
\end{align}
as well as their conjugates. Here, the first two equalities select the lowest scalar component of the $\mathcal{N} = (1,1)$ particle 
multiplet for legs four and five. While the last two relations arise from the integrations over the $\eta_{1,2}/\bar\eta_{1,2}$ 
Grassmann variables that project on the top component of the same multiplet on legs one and two. The third leg is left 
unconstrained. In this manner, we find for the $\Omega$-structure entering Eq.\ \re{CompactA5}
\begin{align}
\Omega_{12345}^{\rm sing} 
=
\frac{S_{45}}{S_{12}} \bra{Q_3} \bar{P}_2 P_1 \bar{P}_5 P_4 | \bar{Q}_3]
\, .
\end{align}
Accounting for an overall Jacobian $S_{12}^2$, which stems from the Grassmann integrations\footnote{For brevity, we do not display these 
integration explicitly anywhere.} of the legs one and two of the supermomentum conserving delta function in Eq.\ \re{GenericAn}, we find
\begin{align}
\label{SingletA5}
\big. \mathcal{A}_5^{(0)} \big|^{\rm sing}
&
=
- \frac{\bra{Q_3} \bar{P}_2 P_1 \bar{P}_5 P_4 | \bar{Q}_3]}{S_{23} S_{34} S_{51}}
\, .
\end{align}
This expression contains both the MHV and $\overline{\rm MHV}$ amplitudes, depending on the choice of remaining Grassmann 
variables for the third leg. For instance, to extract the MHV amplitude $\vev{\phi'''\phi''' g^+ \phi\phi}$, we set 
$\eta_{3,1} = \bar\eta_3^{\dot{2}} = 0$ and find for its four-dimensional projection
\begin{align}
\big. \mathcal{A}_5^{(0)} \big|^{\rm sing}_{\rm MHV}
&
=
\frac{\vev{12}\vev{45}}{\vev{23} \vev{34} \vev{51}} \times \big(- \bar\eta_3^{\dot{1}} \eta_{3,2}\big)
\, ,
\end{align}
upon setting all extra-dimensional (from the four-dimensional point of view) components of the Weyl spinors to zero \cite{Belitsky:2024rwv}.

Generalizing the above consideration, for six legs, we will look for a component corresponding to the four-dimensional MHV amplitude 
$\vev{\phi'''\phi''' g^+ \phi\phi g^+}$. To obtain it, we again set the supercharges for the fourth and fifth legs to zero, and integrate over 
$\eta_{1,2}$ and $\bar\eta_{1,2}$. The latter imposes constraints on the value of $Q_1/\bar{Q}_1$ and $Q_2/\bar{Q}_2$. When assembled 
together, we will dub this as the `singlet' projection as well
\begin{align}
\label{6LegSingletConds}
{\rm sing}
=
\left\{
Q_4 = Q_5 = 0
\, , 
Q_1 = - \frac{Q_{36} \bar{P}_2 P_1}{S_{12}}
\, , 
Q_2 = - \frac{Q_{36} \bar{P}_1 P_2}{S_{12}}
\right\}
\, .
\end{align}
Here $Q_{36} \equiv Q_3 + Q_6$. Identical conditions are implemented on their chiral conjugate values by applying the cc-operation to 
Eq.\ \re{6LegSingletConds}. Then, the $\Omega$'s simplify drastically. We find
\begin{align}
\Omega_i^{\rm sing} = \ft12 \bra{b_i} \bar{b}_{i+1}] + {\rm cc}
\, ,
\end{align}
in terms of just six brackets,
\begin{align}
&
\bra{b_1} 
= - S_{123} \bra{Q_1} - \bra{Q_6} \bar{P}_{23} P_1
\, , \quad
&&
\bra{b_4} 
= \bra{Q_6} \bar{P}_5 P_4
\, , \nonumber\\
&
\bra{b_2} 
= - S_{34} \bra{Q_2} - \bra{Q_3} \bar{P}_4 P_1
\, , \quad
&&
\bra{b_5} 
= \bra{Q_{16}} \bar{P}_{16} P_5
\, , \\
&
\bra{b_3} 
= - S_{45} \bra{Q_3}
\, , \quad
&&
\bra{b_6} 
= - S_{612} \bra{Q_6} - \bra{Q_3} \bar{P}_{12} P_6
\, . \nonumber
\end{align}
These can be further reduced to a set of independent structures
\begin{align}
&
\mathcal{U}_{12} = \bra{Q_3} \bar{P}_2 P_1 \bar{P}_6 P_{12} | \bar{Q}_3 ]
\, , \quad
&&
\mathcal{U}_{23} = \bra{Q_3} \bar{P}_2 P_1 \bar{P}_{23} P_4 | \bar{Q}_3 ]
\, , \\
&
\mathcal{V}_{12} = \bra{Q_6} \bar{P}_{12} P_3 \bar{P}_2 P_1 | \bar{Q}_3 ]
\, , \quad
&&
\mathcal{V}_{16} = \bra{Q_6} \bar{P}_5 P_{61} \bar{P}_2 P_1 | \bar{Q}_6 ]
\, , \\
&
\mathcal{W}_{12} = \bra{Q_3} \bar{P}_2 P_1 | \bar{Q}_6]
\, , \quad
&&
\mathcal{W}_{45} = \bra{Q_3} \bar{P}_4 P_5 | \bar{Q}_6]
\, ,
\end{align}
to read
\begin{align}
\Omega_1^{\rm sing} 
&
= - \frac{S_{123}}{2 S_{12}} \mathcal{U}_{23} - \frac{S_{34}}{2 S_{12}} \mathcal{V}_{12} 
- 
\frac{S_{123} S_{34}}{S_{12}} \mathcal{W}_{12} - \frac{S_{23}}{2} \mathcal{W}_{45} + {\rm cc}
\, , \\
\Omega_2^{\rm sing}
&
= - \frac{S_{45}}{2 S_{12}} \mathcal{U}_{23} - \frac{S_{34} S_{45}}{2 S_{12}} \mathcal{W}_{12}  + {\rm cc}
\, , \\
\Omega_3^{\rm sing}
&
= \frac{S_{45}}{2} \mathcal{W}_{45} + {\rm cc}
\, , \\
\Omega_4^{\rm sing}
&
= - \frac{S_{45}}{2 S_{12}} \mathcal{V}_{16} - \frac{S_{45} S_{56}}{2 S_{12}} \mathcal{W}_{12}  + {\rm cc}
\, , \\
\Omega_5^{\rm sing}
&
= - \frac{S_{56}}{2 S_{12}} \mathcal{U}_{12} - \frac{S_{612}}{2 S_{12}} \mathcal{V}_{16} 
- 
\frac{S_{612} S_{56}}{S_{12}} \mathcal{W}_{12} - \frac{S_{16}}{2} \mathcal{W}_{45} + {\rm cc}
\, , \\
\Omega_6^{\rm sing}
&
= - \frac{S_{123}}{2 S_{12}} \mathcal{U}_{12} - \frac{S_{612}}{2 S_{12}} \mathcal{V}_{12} 
- 
\frac{S_{612} S_{123}}{S_{12}} \mathcal{W}_{12} - \frac{S_{45}}{2} \mathcal{W}_{12} + {\rm cc}
\, .
\end{align}
The numerator of the amplitude then becomes
\begin{align}
\label{N6singlet}
\mathcal{N}^{\rm sing}_6
=
&
S_{45} S_{123} S_{612}
\bigg[
\frac{S_{12}^2 S_{23} S_{61}}{S_{123} S_{612}} \mathcal{W}_{45} \widebar{\mathcal{W}}_{45}
-
S_{34} S_{56} (4 - U_1 - 2 U_2 - 2 U_3) \mathcal{W}_{12} \widebar{\mathcal{W}}_{12}
\nonumber\\
+
&
\frac{S_{12} S_{23} S_{56} U_3}{S_{123}} [\mathcal{W}_{12} \widebar{\mathcal{W}}_{45} + \mathcal{W}_{45} \widebar{\mathcal{W}}_{12}]
-
\frac{S_{34} (1 - U_2)}{4 S_{612}}  [\mathcal{U}_{12} + \widebar{\mathcal{U}}_{12}]  [\mathcal{V}_{16} + \widebar{\mathcal{V}}_{16}]
\nonumber\\
-
&
\frac{S_{56} (1 - U_3)}{4 S_{123}}  [\mathcal{U}_{23} + \widebar{\mathcal{U}}_{23}]  [\mathcal{V}_{12} + \widebar{\mathcal{V}}_{12}]
-
\frac{(2 - U_2 - U_3)}{4} [\mathcal{U}_{23} + \widebar{\mathcal{U}}_{23}]  [\mathcal{V}_{16} + \widebar{\mathcal{V}}_{16}]
\bigg]
\, ,
\end{align}
where we accounted for an overall Jacobian factor $S_{12}^2$ from integrating out legs one and two, and we kept only Grassmann products 
that are nonvanishing for the case of the legs three and six being gluons\footnote{Were this taken to be scalars, the amplitude would correspond
to NMHV upon the dimensional reduction.}. The barred variables are the chiral conjugates of unbarred
\begin{align}
\left( \widebar{\mathcal{U}}, \widebar{\mathcal{V}}, \widebar{\mathcal{W}} \right) \equiv {\rm cc} \left( \mathcal{U}, \mathcal{V}, \mathcal{W} \right)
\, .
\end{align}

It is instructive to observe the cancellation of the spurious pole first in four dimensions. To extract the four-dimensional MHV amplitude 
$\vev{\phi''' \phi''' g^+ \phi\phi g^+}$, we set $\eta_{3,1} = \bar\eta_3^{\dot{2}} = \eta_{6,1} = \bar\eta_6^{\dot{2}} = 0$ and reduce the 
six-dimensional spinors down to four by setting the extra-dimensional momentum components to zero. Then, choosing
$\mathcal{W}_{12} \widebar{\mathcal{W}}_{45}$ as the sole independent structure, we can reduce the rest of the appearing ones 
in Eq.\ \re{N6singlet} to it via the following relations
\begin{alignat}{2}
\label{4DMHVfirst}
\left.\frac{\mathcal{W}_{12} \widebar{\mathcal{W}}_{12}}{\mathcal{W}_{12} \widebar{\mathcal{W}}_{45}}\right|_{\rm MHV}
&
=
- \frac{\vev{12} [23][61]}{[34]\vev{45}[56]} 
&&
= 
\frac{\tr_2 [p_1 p_2 p_3 p_4 p_5 p_6]}{\tr_2 [p_3 p_4 p_5 p_6 p_5 p_4]}
\\
&
&&
= 
\frac{s_{612} s_{123} s_{234}}{2 s_{34} s_{45} s_{56}} (1 - u_1 - u_2 - u_3)
\, , \nonumber\\
\left.\frac{\mathcal{W}_{45} \widebar{\mathcal{W}}_{45}}{\mathcal{W}_{12} \widebar{\mathcal{W}}_{45}}\right|_{\rm MHV}
&
=
- \frac{[34]\vev{45}[56]}{\vev{12}[23][61]} 
&&
= 
\frac{\tr_2 [p_3 p_4 p_5 p_6 p_1 p_2]}{\tr_2 [p_3 p_2 p_1 p_6 p_1 p_2]}
\\
&
&&
=
\frac{s_{612} s_{123} s_{234}}{2 s_{12} s_{23} s_{61}} (1 - u_1 - u_2 - u_3)
\, , \nonumber\\
\left.\frac{\mathcal{U}_{23} \mathcal{V}_{16}}{\mathcal{W}_{12} \widebar{\mathcal{W}}_{45}}\right|_{\rm MHV}
&
=
- \frac{\bra{5} p_{16} p_2 p_1 p_{23} \ket{4}}{\vev{45}} 
&&
= 
\frac{\tr_2 [p_1 p_{23} p_4 p_5 p_{16} p_2]}{\tr_2 [p_4 p_5]}
\\
&
&&
=
- \frac{s_{612} s_{123} s_{234}}{2 s_{45}} (1 + u_1 - u_2 - u_3)
\, , \nonumber\\
\left.\frac{\mathcal{U}_{23} \mathcal{V}_{12}}{\mathcal{W}_{12} \widebar{\mathcal{W}}_{45}}\right|_{\rm MHV}
&
=
\frac{[6|p_{12} p_3 p_2 p_1 p_{23} \ket{4}}{\vev{45} [56]} 
&&
= 
\frac{\tr_2 [p_6 p_{12} p_3 p_2 p_1 p_{23} p_4 p_5]}{\tr_2 [p_5 p_6 p_4 p_4]}
\\
&
&&
=
- \frac{s_{612} s_{123}^2 s_{234}}{2 s_{45} s_{56}} (1 - u_1 - u_2 - u_3 + 2 u_1 u_2)
\, , \nonumber\\
\label{4DMHVlast}
\left.\frac{\mathcal{U}_{12} \mathcal{V}_{16}}{\mathcal{W}_{12} \widebar{\mathcal{W}}_{45}}\right|_{\rm MHV}
&
=
\frac{[5|p_{61} p_2 p_1 p_6 p_{13} \ket{4}}{[34] \vev{45}} 
&&
= 
\frac{\tr_2 [p_5 p_{61} p_2 p_1 p_6 p_{12} p_3 p_4]}{\tr_2 [p_5 p_4 p_3 p_4]}
\\
&
&&
=
- \frac{s_{612}^2 s_{123} s_{234}}{2 s_{34} s_{45}} (1 - u_1 - u_2 - u_3 + 2 u_1 u_3)
\, ,
\end{alignat}
(and similar expressions for structures not shown explicitly). Here, we rationalized the numerators and denominators to two-dimensional 
Dirac strings, which were then in turn computed with {\tt FeynCalc} \cite{Mertig:1990an,Shtabovenko:2023idz}. Assembling these expressions
in Eq.\ \re{N6singlet}, the spurious pole cancels, and we obtain the well-known result
\begin{align}
\
\big. \mathcal{A}_6^{(0)} \big|^{\rm sing}_{\rm MHV}
&
=
- 
\left.
\frac{\mathcal{W}_{12} \widebar{\mathcal{W}}_{45} + \mathcal{W}_{45} \widebar{\mathcal{W}}_{12}}{2 S_{23} S_{34} S_{56} S_{61}}
\right|_{\rm MHV}
\\
&=
\frac{\vev{12}\vev{45}}{\vev{23} \vev{34} \vev{56} \vev{61}} \times \big(- \bar\eta_3^{\dot{1}} \eta_{32} \bar\eta_6^{\dot{1}} \eta_{62}\big)
\, . \nonumber
\end{align}
Here both structures possess the same four-dimensional MHV projection, $\mathcal{W}_{12} \widebar{\mathcal{W}}_{45}|_{\rm MHV} = 
\mathcal{W}_{45} \widebar{\mathcal{W}}_{12}|_{\rm MHV}$.

Returning to six dimensions, from now on, we are only interested in the overlap with a structure, which reduces to the above MHV amplitude. 
In all calculations that follow, gluons will solely appear in external states, so this will not be much of a limitation. Any chiral-even structure 
appearing in Eq.\ \re{N6singlet} will do the job, but for simplicity reasons and a more transparent four-dimensional limit that we have already 
worked out above, we select
\begin{align}
\label{W12Wb45structure}
\mathcal{W}_{12} \widebar{\mathcal{W}}_{45} + \mathcal{W}_{45} \widebar{\mathcal{W}}_{12}
\, .
\end{align}
To implement the extraction in a concise fashion, we choose to do it by preserving {\sl six-dimensional covariance}, on the one hand, 
as well as to maintain its nice dual conformal inversion properties, on the other hand. This is done with the following projector
\begin{align}
\label{Pi36projector}
\Pi_{36} = \bra{Q_3} Z \bar{P}_2 P_1 |\bar{Q}_6] \bra{Q_6} Z \bar{P}_5 P_4 | \bar{Q}_3]
\, .
\end{align}
However, it undiscriminately accounts for both chiral and antichiral degrees of freedom, contributing equally to Dirac traces since 
$\mathcal{N} = (1,1)$ is a non-chiral sYM. When the Grassmann integration yields the same number of these, whose ratio defines 
the relative coefficient we are after, this will not be a problem, as the number of contributing degrees of freedom cancels between the 
two. However, when it differs\footnote{Notice that in four dimensions, the product of traces of Dirac strings of momenta can be recoupled 
into a single one when some of the repeating massless momenta occur in pairs. We do not have this luxury in six dimensions since the 
little group is non-Abelian and massless six-dimensional momenta are defined by a {\sl sum} of products of pairs of Weyl spinors 
\re{SHmomentum} rather than just a single tensor product $p = \ket{p} [p|$.}, there will be an over- or under-counting, depending on 
the perspective. Thus, to avoid it, we introduced a wildcard $Z$-marker in Eq.\ \re{Pi36projector}. When two $Z$'s belong to the same 
trace $Z^2 \to Z$, and the trace with just one is understood as
\begin{align}
\tr_4 [Z \dots] = \ft12 \tr_4[\dots]
\, .
\end{align}
This affects only quantities involving different relative number of traces. Introducing
\begin{align}
\label{DoubleBrackets}
\vev{\vev{\dots}} \equiv \int d^2 \eta_3 d^2 \bar\eta_3 d^2 \eta_6 d^2 \bar\eta_6  (\dots) \Pi_{36}
\, , 
\end{align}
we find for the quotients to the desired independent structure in Eq.\ \re{W12Wb45structure}
\begin{align}
\frac{\vev{\vev{\mathcal{W}_{12} \widebar{\mathcal{W}}_{12}}}}{\vev{\vev{\mathcal{W}_{12} \widebar{\mathcal{W}}_{45}}}}
&=
\frac{\tr_4[P_6 \bar{P}_5 P_4 \bar{P}_3 P_2 \bar{P}_1]}{\tr_4[P_6 \bar{P}_5 P_4 \bar{P}_3 P_4 \bar{P}_5]}
\nonumber\\
&
= \frac{S_{612} S_{123} S_{234}}{2 S_{34} S_{45} S_{56}} (1 - U_1 - U_2 - U_3)
\, , \\
\frac{\vev{\vev{\mathcal{W}_{45} \widebar{\mathcal{W}}_{45}}}}{\vev{\vev{\mathcal{W}_{12} \widebar{\mathcal{W}}_{45}}}}
&=
\frac{\tr_4[P_6 \bar{P}_1 P_2 \bar{P}_3 P_4 \bar{P}_5]}{\tr_4[P_6 \bar{P}_1 P_2 \bar{P}_3 P_2 \bar{P}_1]}
\nonumber\\
&
= \frac{S_{612} S_{123} S_{234}}{2 S_{12} S_{23} S_{61}} (1 - U_1 - U_2 - U_3)
\, , \\
\frac{\vev{\vev{\mathcal{U}_{23} \mathcal{V}_{16}}}}{\vev{\vev{\mathcal{W}_{12} \widebar{\mathcal{W}}_{45}}}}
&=
\frac{\tr_4[P_6 \bar{P}_1 P_2 \bar{P}_3 P_2 \bar{P}_1 P_{23} \bar{P}_4 P_3 \bar{P}_4 P_5 \bar{P}_6 
P_5 \bar{P}_{16} P_2 \bar{P}_1]
}{
\ft12 \tr_4[P_6 \bar{P}_5 P_4 \bar{P}_3 P_4 \bar{P}_5] \tr_4[P_6 \bar{P}_1 P_2 \bar{P}_3 P_2 \bar{P}_1]}
\nonumber\\
&
= - \frac{S_{612} S_{123} S_{234}}{2 S_{45}} (1 + U_1 - U_2 - U_3)
\, , \\
\frac{\vev{\vev{\mathcal{U}_{23} \mathcal{V}_{12}}}}{\vev{\vev{\mathcal{W}_{12} \widebar{\mathcal{W}}_{45}}}}
&=
\frac{\tr_4[P_6 \bar{P}_1 P_2 \bar{P}_3 P_2 \bar{P}_1 P_{23} \bar{P}_4 P_3 \bar{P}_4 P_5 \bar{P}_6 P_{12} \bar{P}_3 P_{2} \bar{P}_1]
}{
\ft12 \tr_4[P_6 \bar{P}_5 P_4 \bar{P}_3 P_4 \bar{P}_5] \tr_4[P_6 \bar{P}_1 P_2 \bar{P}_3 P_2 \bar{P}_1]}
\nonumber\\
&
=
- \frac{S_{612} S_{123}^2 S_{234}}{2 S_{45} S_{56}} (1 - U_1 - U_2 - U_3 + 2 U_1 U_2)
\, , \\
\frac{\vev{\vev{\mathcal{U}_{12} \mathcal{V}_{16}}}}{\vev{\vev{\mathcal{W}_{12} \widebar{\mathcal{W}}_{45}}}}
&=
\frac{\tr_4[P_6 \bar{P}_1 P_2 \bar{P}_3 P_2 \bar{P}_1 P_6 \bar{P}_{12} P_3 \bar{P}_4 P_5 \bar{P}_6 P_5 \bar{P}_{16} P_{2} \bar{P}_1]
}{
\ft12 \tr_4[P_6 \bar{P}_5 P_4 \bar{P}_3 P_4 \bar{P}_5] \tr_4[P_6 \bar{P}_1 P_2 \bar{P}_3 P_2 \bar{P}_1]}
\nonumber\\
&
=
- \frac{S_{612}^2 S_{123} S_{234}}{2 S_{34} S_{45}} (1 - U_1 - U_2 - U_3 + 2 U_1 U_3)
\, .
\end{align} 
Here we used
\begin{align}
\label{IntQQ2P}
\int d^2 \eta_i \ket{Q_i} \bra{Q_i} = - P_i
\, ,
\end{align}
(as well as its cc) and calculated the resulting traces with {\tt FeynCalc} \cite{Mertig:1990an,Shtabovenko:2023idz}. Notice that the
six-dimensional parity-odd contribution, see Appendix \ref{6DAppendix} vanishes by momentum conservation. The other chiral half is 
found in a similar manner with a chiral conjugate projector $\bar{\Pi}_{36}$. Their respective coefficients are, of course, identical to the above. 
It is clear that they are just a six-dimensional uplift, $s_{ij\dots} \to S_{ij\dots}$, of the four-dimensional ones \re{4DMHVfirst}--\re{4DMHVlast}. 
The six-dimensional `singlet' component of the six-leg amplitude that takes centre stage in our construction below is then
\begin{align}
\label{treeA6}
\big. \mathcal{A}_{6}^{(0)} \big|^{\rm sing}
=
- 
\frac{\mathcal{W}_{12} \widebar{\mathcal{W}}_{45} + \mathcal{W}_{45} \widebar{\mathcal{W}}_{12}}{2 S_{23} S_{34} S_{56} S_{61}}
\, .
\end{align}

Before closing this section, let us offer a few comments regarding the drawbacks of our extraction procedure, which we nevertheless
favor, being ultimately interested in the off-shell kinematics of external states from the four-dimensional point of view. This implies that
we need to preserve six-dimensional covariance. First, we will be insensitive to the so-called $\mu$-terms \cite{Bern:2002tk}. In order 
to extract them, one has to perform calculations in a noncovariant fashion by splitting the spinor algebra in four- and 
out-of-four-dimensional, i.e., mass, components as was done in Ref.\ \cite{Bern:2010qa}. In the present two-loop context, this 
appears to lead to very cumbersome algebra. We favor efficiency over completeness: the $\mu$-terms are known to merely 
generate $O (\ep)$ in the logarithm of the amplitude in the massless case, and would be irrelevant for the special Coulomb branch 
since the loop momenta are restricted to the four-dimensional space-time anyway. Second, the four-dimensional parity-odd contributions 
cancel between chiral and antichiral degrees of freedom in all traces involved, and, therefore, they will be `missed' by our projection
as well. Again, at the conformal point, they vanish as $\varepsilon \to 0$ after exponentiation \cite{Cachazo:2008hp}.

\section{Six-dimensional loops}
\label{LoopSection}

The main goal of this paper is to determine a set of integrals that define quantum corrections to \re{LessSymA6} at one- and two-loop
orders. This corresponds to the Coulomb-branch generalization of the MHV amplitude. The analogue of the NMHV amplitude is presently 
not addressed and is left as a problem for the future. With this stipulation in mind, we will drop the superscript `sing' that specifies the 
component used for this purpose. The perturbative expansion of the amplitude in the six-dimensional 't Hooft coupling 
$\lambda_6 = g_{\rm\scriptscriptstyle 6, YM}^2 N_c$ reads up to the second order
\begin{align}
\mathcal{A}_{6} 
=
\mathcal{A}^{(0)}_{6}
+ 
\lambda_6 \, \mathcal{A}^{(1)}_{6}
+ 
\lambda_6^2 \, \mathcal{A}^{(2)}_{6} + \dots
\, .
\end{align}
Here, the one- and two-loop amplitudes will be determined, after pulling out the tree amplitude, as linear combinations of six-dimensional momentum 
integrals $\mathcal{I}^{(\ell)}_\alpha$ accompanied by polynomials in two- and three-particle Mandelstam invariants $c_\alpha^{(\ell)}$,
\begin{align}
\mathcal{A}^{(\ell)}_{6} = \mathcal{A}^{(0)}_{6}  \sum_\alpha c^{(\ell)}_\alpha \mathcal{I}^{(\ell)}_\alpha
\, .
\end{align}
The special Coulomb branch is reached by passing to the dual bosonic coordinates $X_i = (x_i, y_i)$ and compactifying the six-dimensional 
 {\sl loop integrals} on a two-torus of equal radii $R_0$ in $y$-space. This reduction yields an overall factor of $(2 \pi R_0)^2$ 
accompanying the dimensionally reduced element of the four-volume $d^4 x_0$, 
\begin{align}
\label{DRmeasure}
d^6 X_0 \to (2 \pi R_0)^2 \delta^{(2)} (y_0) d^4 x_0
\, .
\end{align}
This sets our convention of the per-loop measure to be
\begin{align}
\label{6Dto4D}
\lambda_6 \int \frac{d^6 X_0}{i (2\pi)^6} 
\ 
\stackrel{\rm DR}{\longrightarrow}
\ 
g^2 \int \frac{d^4 x_0}{i \pi^2}
\, ,
\end{align}
with $g^2 = R_0^2 \lambda_6/(4 \pi)^2$ being the four-dimensional 't Hooft coupling. Effectively, this corresponds to setting the 
out-of-four-dimensional components $y_0$ and $y_{0^\prime}$ of the dual variables to zero. The extra-dimensional components of 
external legs are kept at generic values, however. These are the off-shellness parameters we referred to earlier.

\subsection{One-loop integrand}
\label{1LoopSection}

Now we turn to the unitarity-cut sewing technique to construct loop integrands. Improved ultraviolet properties of the maximally 
supersymmetric Yang-Mills theory imply the absence of subgraphs with bubbles and triangles. We start at one loop, where a 
single cut carrying a three-particle invariant suffices to fix it unambiguously \cite{Bern:1994zx,Bern:1994cg}. For the case at hand, 
it is shown in the right panel of Fig.\ \ref{TreeOneLoopPic}. It reads
\begin{align}
\label{1loopCut}
\big. \mathcal{A}_6^{(1)} \big|_{S_{123}-{\rm cut}}
&
=
\delta^{(4)} \left( Q_{123456} \right)
\delta^{(4)} \left( \bar{Q}_{123456} \right)
\int \prod_{i=1}^2 d^2 \eta_{K_i} d^2 \bar\eta_{K_i}
\delta^{(4)} \left( Q_{456 K_1 K_2} \right)
\delta^{(4)} \left( \bar{Q}_{456 K_1 K_2} \right)
\nonumber\\
&
\times
\widehat{\mathcal{A}}_5^{(0)} (P_1, P_2, P_3, - K_1, - K_2)
\widehat{\mathcal{A}}_5^{(0)} (K_1, K_2, P_6, P_5, P_4)
\, ,
\end{align}
where we combined the arguments of the Grassmann delta functions of each contributing five-leg amplitude, so that we can pull 
out the overall supercharge conservation.

To determine the integrand, we will now impose the singlet channel conditions \re{6LegSingletConds}, which enforce $Q_4 = Q_5 = 0$,
and its chiral conjugate, as well as eliminates the overall supermomentum delta functions in Eq.\ \re{1loopCut}, leaving just the 
Jacobian $S_{12}^2$,
\begin{align}
\big. \mathcal{A}_6^{(1)} \big|^{\rm sing}_{S_{123}-{\rm cut}}
&
=
\int \prod_{i=1}^2 d^2 \eta_{K_i} d^2 \bar\eta_{K_i}
\delta^{(4)} \left( Q_{6 K_1 K_2} \right)
\delta^{(4)} \left( \bar{Q}_{6 K_1 K_2} \right)
\nonumber\\
&
\times
S_{12}^2
\widehat{\mathcal{A}}_5^{(0)} (P_1, P_2, P_3, - K_1, - K_2) |_{{\rm Super}'_1}
\widehat{\mathcal{A}}_5^{(0)} (K_1, K_2, P_6, P_5, P_4) |_{{\rm Super}_2}
\, .
\end{align}
The leftover supercharge conditions on the involved amplitudes are
\begin{align}
{\rm Super}'_1
&
=
\left\{
Q_1 = - \frac{Q_{36} \bar{P}_2 P_1}{S_{12}} , Q_2 = - \frac{Q_{36} \bar{P}_1 P_2}{S_{12}}
\right\}
\, , \\
{\rm Super}_2
&
=
\left\{
Q_4 = Q_5 = 0, Q_{K_1} = - \frac{Q_6 \bar{K}_2 K_1}{S_{K_1 K_2}} , Q_{K_2} = - \frac{Q_6 \bar{K}_1 K_2}{S_{K_1 K_2}}
\right\}
\, .
\end{align}
Their chiral conjugates are tacitly implied as well. The second amplitude in the above expression is then nothing but Eq.\ \re{SingletA5}
with an overall Jacobian missing
\begin{align}
\label{SecondA5}
\widehat{\mathcal{A}}_5^{(0)} (K_1, K_2, P_6, P_5, P_4) |_{{\rm Super}_2}
=
- \frac{\bra{Q_6} \bar{K}_2 K_1 \bar{P}_4 P_5 | \bar{Q}_6]}{S_{K_1 K_2}^2 S_{4 K_1} S_{6 K_2} S_{56}}
\, .
\end{align}

%%%%%%%%%%%%%%%%%%%%%%%%%%%%%%%%%%%%%%%%%%%%%%%%%%%%%%%%%%%%%%%%%%%%%
%            Figure
%%%%%%%%%%%%%%%%%%%%%%%%%%%%%%%%%%%%%%%%%%%%%%%%%%%%%%%%%%%%%%%%%%%%%
\begin{figure}[t]
\begin{center}
\mbox{
\begin{picture}(0,90)(120,0)
\put(0,0){\insertfig{8}{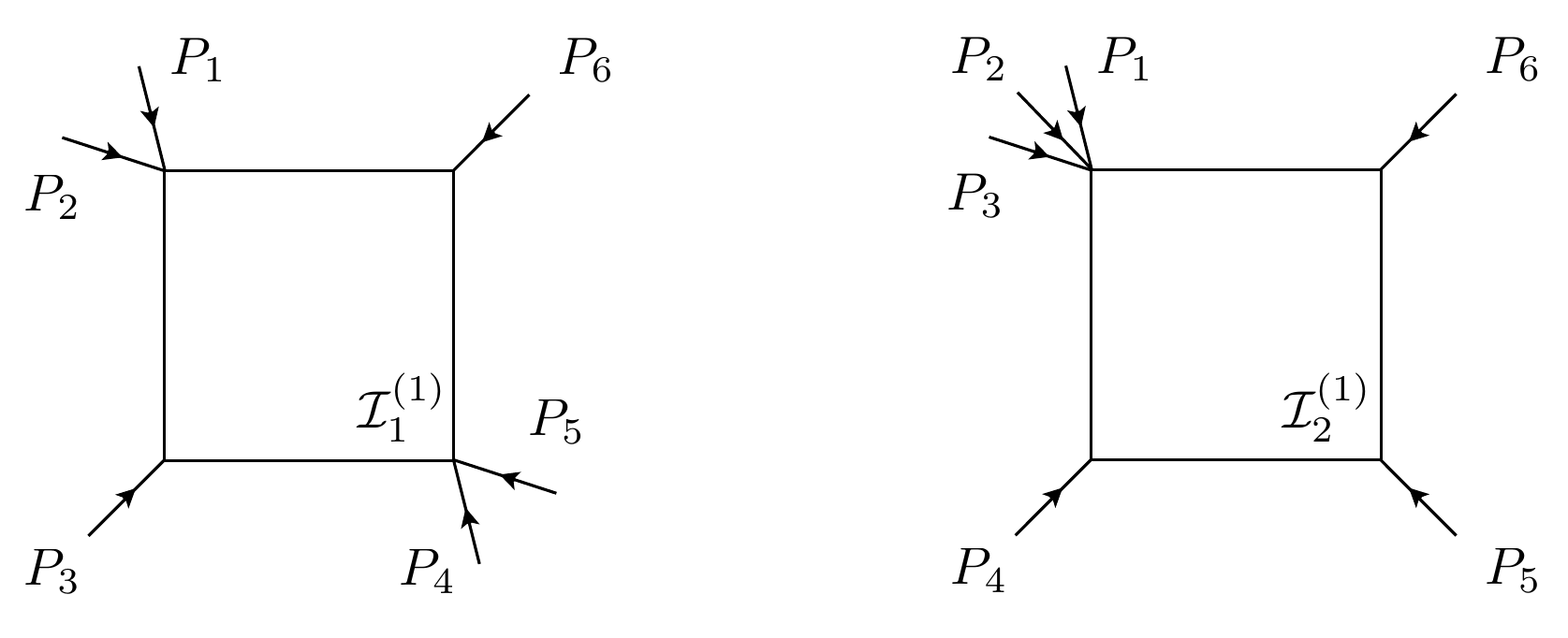}}
\end{picture}
}
\end{center}
\caption{\label{Ints1L} Basis one-loop integrals for $\mathcal{A}_6^{(1)}$ in Eq.\ \re{1loopA6}.}
\end{figure}
%%%%%%%%%%%%%%%%%%%%%%%%%%%%%%%%%%%%%%%%%%%%%%%%%%%%%%%%%%%%%%%%%%%%%

We want to find the overlap of \re{1loopCut} with the six-leg amplitude in the channel projected by $\Pi_{36}$ given in Eq.\ \re{Pi36projector}.
It will suffice for this purpose to use the less symmetric form of the six-leg amplitude
\begin{align}
\label{LessSymA6}
\big. \mathcal{A}_{6}^{(0)} \big|^{\rm sing}
=
- 
\frac{\mathcal{W}_{12} \widebar{\mathcal{W}}_{45}
}{
S_{23} S_{34} S_{56} S_{61}
}
\, ,
\end{align}
easily accessible with the projector $\Pi_{36}$. Then one immediately observes that the $\eta_6/\bar\eta_6$ integrations in \re{DoubleBrackets} 
are saturated by the explicit $Q_6/\bar{Q}_6$-dependence of \re{SecondA5}, so that $Q_6/\bar{Q}_6$ can be set to zero in every other place in 
the integrand of \re{1loopCut}. In particular, this implies that the Grassmann deltas reduce to $\delta^{(4)} \left( Q_{K_1 K_2} \right) \delta^{(4)} 
\left( \bar{Q}_{K_1 K_2} \right)$ and thus project on the scalar intermediate state only. These enhance the first condition to 
\begin{align}
{\rm Super}'_1
\to 
{\rm Super}_1
&
=
\left\{
Q_{K_1} = Q_{K_2} = 0
\, ,
Q_1 = - \frac{Q_3 \bar{P}_2 P_1}{S_{12}} , Q_2 = - \frac{Q_3 \bar{P}_1 P_2}{S_{12}}
\right\}
\, , 
\end{align}
such that
\begin{align}
\label{FirstA5}
\widehat{\mathcal{A}}_5^{(0)} (P_1, P_2, P_3, - K_1, - K_2) |_{{\rm Super}_1}
=
- \frac{\bra{Q_3} \bar{P}_2 P_1 \bar{K}_2 K_1 | \bar{Q}_3]}{S_{12}^2 S_{23} S_{3,-K_1} S_{1,-K_2}}
\, .
\end{align}
Trivially performing the remaining Grassmann integrations in Eq.\ \re{1loopCut}, which result in the Jacobian $S_{K_1 K_2}^2$, we deduce
for the projection in question
\begin{align}
\VEV{\VEV{
\big. \mathcal{A}_6^{(1)} \big|^{\rm sing}_{S_{123}-{\rm cut}}
}}
&
=
\VEV{\VEV{
\frac{\bra{Q_3} \bar{P}_2 P_1 \bar{K}_2 K_1 | \bar{Q}_3]}{S_{23} S_{3,-K_1} S_{1,-K_2}}
\frac{\bra{Q_6} \bar{K}_2 K_1 \bar{P}_4 P_5 | \bar{Q}_6]}{S_{6 K_2} S_{4 K_1} S_{56}}
}}
\nonumber\\
&
= 
\ft12 \tr_4 [\bar{P}_6 P_1 \bar{K}_2 K_1 \bar{P}_3 P_4 \bar{K}_1 K_2]
\frac{
S_{12} S_{45}
}{
S_{3,-K_1} S_{1,-K_2} S_{4 K_1} S_{6 K_2}
}
\, .
\end{align}
Comparing this to the projection of Eq.\ \re{LessSymA6},
\begin{align}
\VEV{\VEV{\big. \mathcal{A}_{6}^{(0)} \big|^{\rm sing}}}
=
S_{12} S_{45}
\, ,
\end{align}
we find the integrand (after restoring the cut propagators) in terms of six-dimensional hexagons \cite{Bern:1994zx},
\begin{align}
{\rm Integrand}^{(1)}_{S_{123}}
=
\frac{
\ft12 \tr_4 [\bar{P}_6 P_1 \bar{K}_2 K_1 \bar{P}_3 P_4 \bar{K}_1 K_2]
}{
K_1^2 K_2^2 S_{3,-K_1} S_{1,-K_2} S_{4 K_1} S_{6 K_2}
}
\, ,
\end{align}
where $K_1 + K_2 = P_1 + P_2 + P_3 = - P_4 - P_5 - P_6$. The lower label in the ${\rm Integrand}^{(1)}$ refers to its origin
in the cut-sewing procedure. This expression can be further reduced to the sum of boxes with loop-momentum-dependent 
numerators,
\begin{align}
\label{1loopIntegrand}
{\rm Integrand}^{(1)}_{S_{123}}
=
&
-
\frac{
\ft12 \tr_4 [P_6 \bar{K}_2 P_3 \bar{K}_1]
}{
K_1^2 K_2^2 S_{3,-K_1} S_{6 K_2}
}
-
\frac{
\ft12 \tr_4 [P_6 \bar{K}_2 P_4 \bar{K}_1]
}{
K_1^2 K_2^2 S_{4 K_1} S_{6 K_2}
}
\nonumber\\
&
-
\frac{
\ft12 \tr_4 [P_1 \bar{K}_2 P_4 \bar{K}_1]
}{
K_1^2 K_2^2 S_{1,-K_2} S_{4 K_1}
}
-
\frac{
\ft12 \tr_4 [P_1 \bar{K}_2 P_3 \bar{K}_1]
}{
K_1^2 K_2^2 S_{3,-K_1} S_{1,-K_2} 
}
\, .
\end{align}
The remaining traces are decomposed in terms of the accompanying denominators and Mandelstam invariants, 
\begin{align}
\tr_4 [P_6 \bar{K}_2 P_3 \bar{K}_1]
&
=
- 2 S_{3,-K_1} S_{6 K_2} 
\\
&
- (S_{123} - S_{45}) S_{3,- K_1}  - (S_{123} - S_{12}) S_{6 K_2} + S_{123} S_{612} - S_{12} S_{45}
\, , \nonumber\\
\tr_4 [P_6 \bar{K}_2 P_4 \bar{K}_1]
&
=
2 S_{4 K_1} S_{6, K_2}  
\\
&
+ (S_{123} - S_{45}) S_{4 K_1} + (S_{123} - S_{56}) S_{6, K_2} + S_{45} S_{56} 
\, , \nonumber\\
\tr_4 [P_1 \bar{K}_2 P_4 \bar{K}_1]
&
=
- 2 S_{1,-K_2} S_{4 K_1} 
\\
&
- (S_{123} - S_{56}) S_{1,-K_2} - (S_{123} - S_{23}) S_{4 K_1} + S_{123} S_{234} - S_{23} S_{56}
\, , \nonumber\\
\tr_4 [P_1 \bar{K}_2 P_3 \bar{K}_1]
&
=
2 S_{3,-K_1} S_{1,-K_2} 
\\
&
+ (S_{123} - S_{23}) S_{3,- K_1} + (S_{123} - S_{12}) S_{1, - K_2} + S_{12} S_{23} 
\, . \nonumber
\end{align}
Here, the products of loop-momentum-dependent Lorentz invariants in the first terms after the equality signs eliminate corresponding 
propagators and induce bubbles, which, however, cancel pairwise in Eq.\ \re{1loopIntegrand}. So do the next two terms as well, 
corresponding to triangles. This is the famed consequence of the improved ultraviolet behavior of the maximally supersymmetric 
Yang-Mills theory. The remaining ones are the loop-momentum-independent invariants, which are the coefficients accompanying 
the two-mass easy and one-mass six-dimensional boxes,
\begin{align}
{\rm Integrand}^{(1)}_{S_{123}}
=
&
- 
\frac{1}{2}
\frac{
S_{123} S_{612} - S_{12} S_{45}
}{
K_1^2 K_2^2 S_{3,-K_1} S_{6 K_2}
}
- 
\frac{1}{2}
\frac{S_{123} S_{234} - S_{23} S_{56}
}{
K_1^2 K_2^2 S_{1,-K_2} S_{4 K_1}
}
\nonumber\\
&
-
\frac{1}{2}
\frac{S_{45} S_{56}
}{
K_1^2 K_2^2 S_{4 K_1} S_{6 K_2}
}
-
\frac{1}{2}
\frac{S_{12} S_{23}
}{
K_1^2 K_2^2 S_{3,-K_1} S_{1,-K_2} 
}
\, .
\end{align}

%%%%%%%%%%%%%%%%%%%%%%%%%%%%%%%%%%%%%%%%%%%%%%%%%%%%%%%%%%%%%%%%%%%%%
%            Figure
%%%%%%%%%%%%%%%%%%%%%%%%%%%%%%%%%%%%%%%%%%%%%%%%%%%%%%%%%%%%%%%%%%%%%
\begin{figure}[t]
\begin{center}
\mbox{
\begin{picture}(0,260)(215,0)
\put(0,0){\insertfig{15}{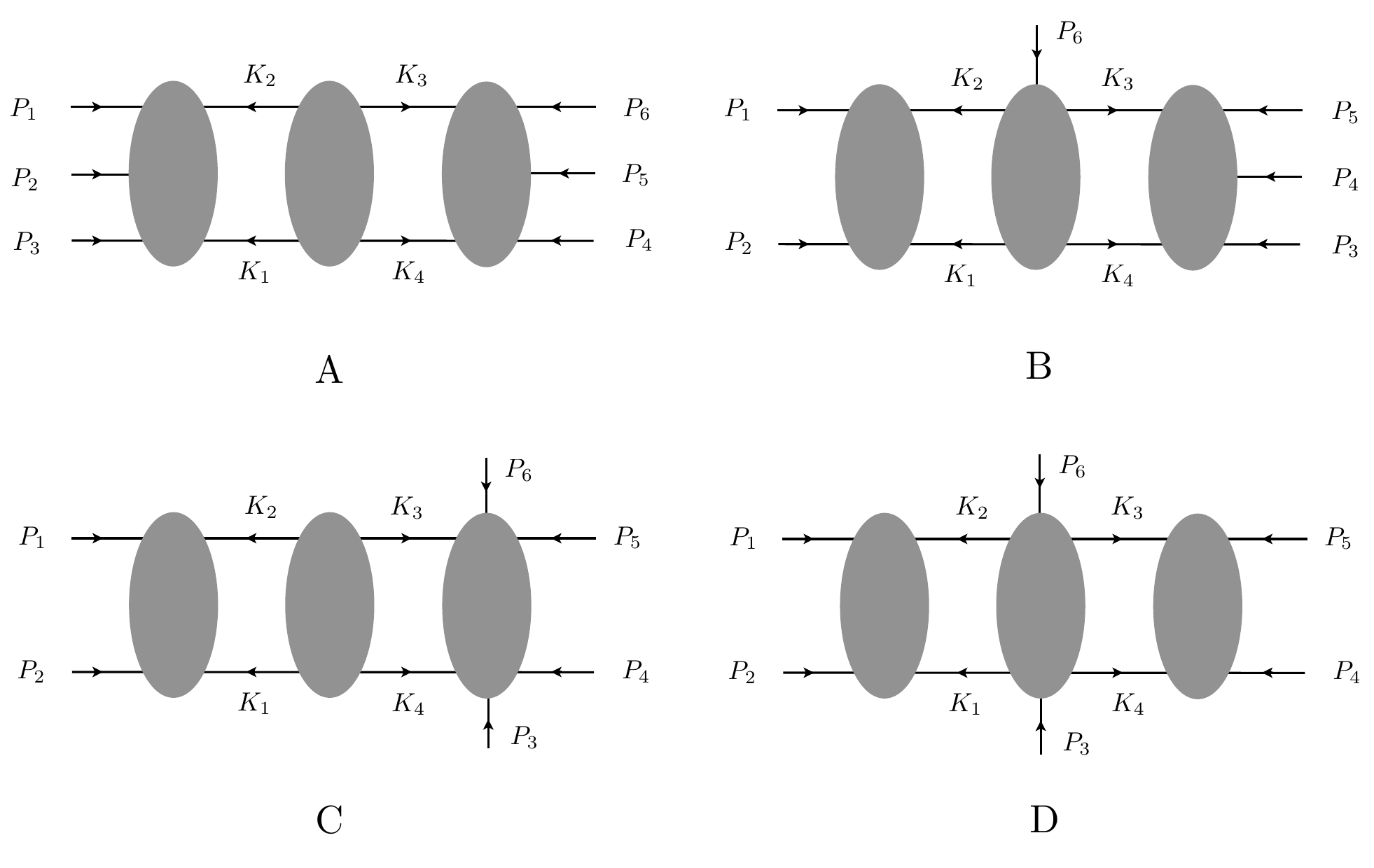}}
\end{picture}
}
\end{center}
\caption{\label{TwoLoopPic} Iterated unitarity cuts that determine the two-loop integrand of the six-leg amplitude.}
\end{figure}
%%%%%%%%%%%%%%%%%%%%%%%%%%%%%%%%%%%%%%%%%%%%%%%%%%%%%%%%%%%%%%%%%%%%%

Of course, different cuts detect some of the same integrals if these possess corresponding nonvanishing discontinuities. This is what we 
observe for the above $S_{123}$-cut. If we choose the first and third terms as the independent contributions to the integrand, the other 
two are just their cyclic images. They emerge as primary contributions in other cuts. Thus, considerations of the other five cuts with 
three-particle invariants will detect all independent terms in the integrand. Equivalently, these can merely be obtained from the above 
two by cyclic permutations\footnote{This is the case since we are only after the six-dimensional amplitude which corresponds to the MHV one
upon dimensional reduction. Had we chosen to calculate the six-dimensional analogue of the NMHV amplitude, we would have to consider 
all channels independently \cite{Kosower:2010yk}. The analysis would be far more complex since the choice of the singlet component would 
not offer any simplifications and would ensure just one intermediate state in all cuts. All on-shell states of the supermultiplet would contribute 
in this case.}. In this manner, we find the one-loop amplitude as a sum of six orderings
\begin{align}
\label{CyclicPerms}
\sigma_6 = \{ (123456),(234561),(345612),(456123),(561234),(612345) \}
\end{align}
of the integrals detected above
\begin{align}
\label{1loopA6}
\mathcal{A}^{(1)}_{6} = \ft12 \mathcal{A}^{(0)}_{6}  \sum_{\sigma_6} 
\left[\ft{1}{2} c^{(1)}_1 \mathcal{I}^{(1)}_1 + c^{(1)}_2 \mathcal{I}^{(1)}_2 \right]
\end{align}
with
\begin{align}
c^{(1)}_1 = 
S_{23} S_{56} - S_{123} S_{234}
\, , \qquad
c^{(1)}_2 = - S_{45} S_{56}
\, .
\end{align}
An additional factor of 1/2 in front of $c^{(1)}_1$ eliminates the double counting from implementing permutations, as there are just three
distinct two-mass easy box integrals. The graphical representation for $\mathcal{I}^{(1)}_\alpha$'s is shown in Fig.\ \ref{Ints1L}. Here and 
below, it is understood that these diagrams represent the propagator structure (and irreducible scalar products, when present) of the 
integrand, accompanied by the measure \re{6Dto4D} but with stripped four-dimensional 't Hooft coupling $g^2$. Massless reduction
of Eq.\ \re{1loopA6} agrees with the well-known one-loop MHV amplitude \cite{Bern:1996ja}.

\subsection{Two-loop integrand}
\label{2LoopSection}

Now, we proceed to two loops. Again, given the no-triangle condition, iterated two-particle cuts are all we need to study in order to fix 
the integrand \cite{Bern:2008ap}. The set of required cuts is shown in Fig.\ \ref{TwoLoopPic}. It will be instructive to compare these with 
the corresponding four-dimensional MHV algebra. Thus, we devote Appendix \ref{Appendix4Dcuts} to recall their expressions.

\subsubsection{Graph A}

\begin{figure}[t]
\begin{center}
\mbox{
\begin{picture}(0,295)(239,0)
\put(0,0){\insertfig{16.8}{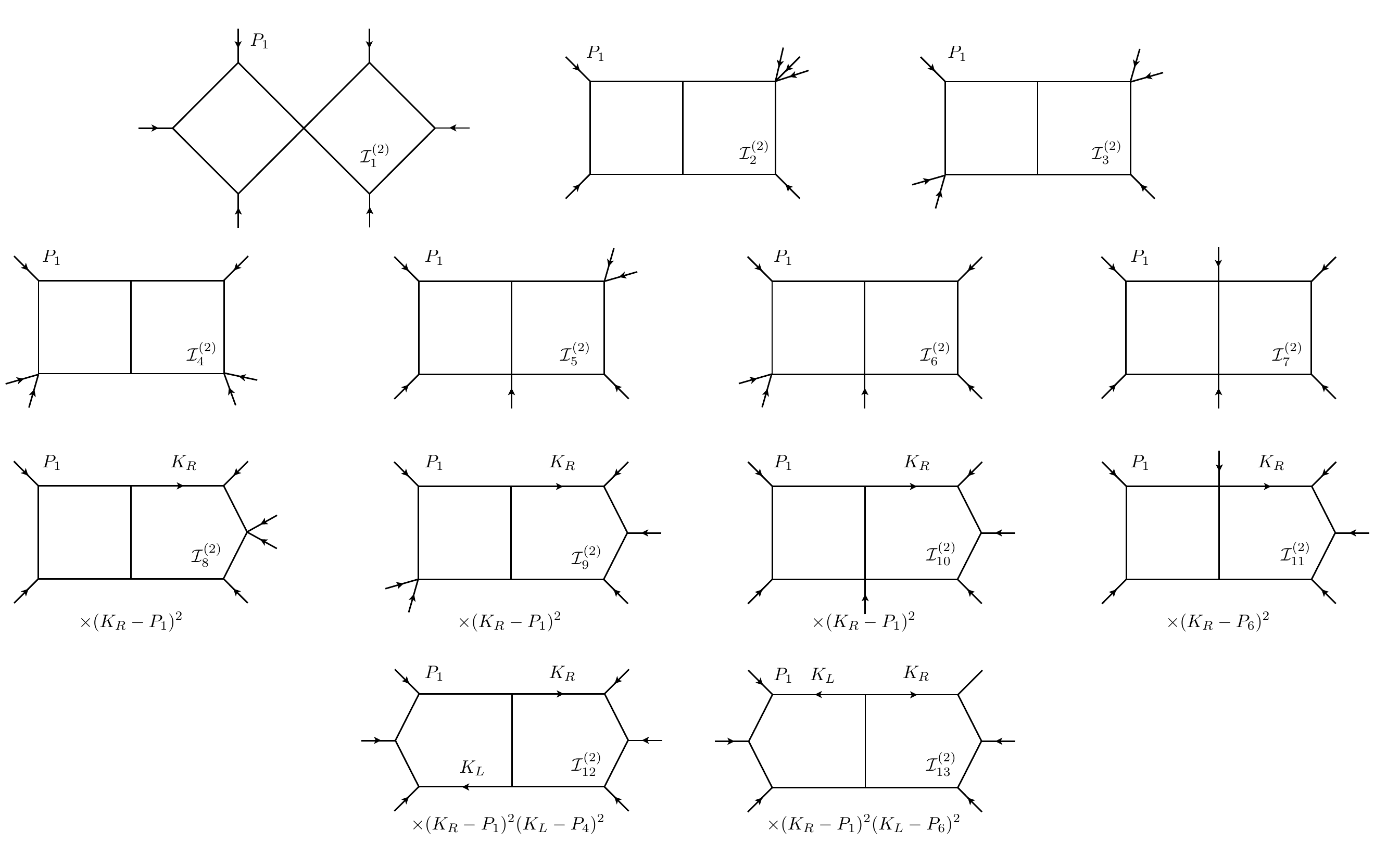}}
\end{picture}
}
\end{center}
\caption{\label{TwoLoopIntergalsPic} The basis of two-loop integrals arising from iterative cuts A, B, C, and D in Fig.\ \ref{TwoLoopPic}. 
As in all other figures, the legs are enumerated counterclockwise. Irreducible numerators are shown explicitly for the chosen routing of 
(left and right) loop momenta $K_L$ and $K_R$.}
\end{figure}

The double cut A in Fig.\ \ref{TwoLoopPic} is an iteration of the one-loop cut considered in the previous section, with the 
intermediate four-leg amplitude \re{A4},
\begin{align}
\label{2loopCuta}
\big. \mathcal{A}_6^{(2)} \big|_{{\rm dcut-A}}
&
=
\delta^{(4)} \left( Q_{123456} \right)
\delta^{(4)} \left( \bar{Q}_{123456} \right)
\int \prod_{i=1}^4 d^2 \eta_{K_i} d^2 \bar\eta_{K_i}
\\
&\times
\delta^{(4)} \left( Q_{K_1 K_2 K_3 K_4} \right)
\delta^{(4)} \left( \bar{Q}_{K_1 K_2 K_3 K_4} \right)
\delta^{(4)} \left( Q_{456 K_3 K_4} \right)
\delta^{(4)} \left( \bar{Q}_{456 K_3 K_4} \right)
\nonumber\\[2mm]
&
\times
\widehat{\mathcal{A}}_5^{(0)} (P_1, P_2, P_3, K_1, K_2)
\widehat{\mathcal{A}}_4^{(0)} (-K_1, - K_2,-K_3, - K_4)
\widehat{\mathcal{A}}_5^{(0)} (K_4, K_3, P_6, P_5, P_4)
\, . \nonumber
\end{align}
So, modifications from the previous calculation are minimal. Namely, imposing the very same singlet condition \re{6LegSingletConds} 
for the external legs, it has the ripple effect of selecting only scalars to propagate across both cuts. With the four-leg amplitude \re{A4} 
being as simple as it is, its denominator $1/[S_{K_1 K_2} S_{K_2 K_3}]$ gets multiplied by the Jacobian $S_{K_1K_2}^2$ from the 
$\eta_{K_1, K_2}/\bar\eta_{K_1, K_2}$ integrations, yielding the sought-after projection
\begin{align}
\VEV{\VEV{
\big. \mathcal{A}_6^{(1)} \big|^{\rm sing}_{{\rm dcut-A}}
}}
&
=
\VEV{\VEV{
\frac{\bra{Q_3} \bar{P}_2 P_1 \bar{K}_2 K_1 | \bar{Q}_3]}{S_{23} S_{1K_2} S_{3K_1}}
\frac{S_{123}}{S_{K_2K_3}}
\frac{\bra{Q_6} \bar{K}_3 K_4 \bar{P}_4 P_5 | \bar{Q}_6]}{S_{6 K_3} S_{4 K_4} S_{56}}
}}
\, ,
\end{align}
where we used $S_{K_1 K_2} = S_{123}$ owing to momentum conservation. Then the integrand is
\begin{align}
\label{2LoopIntegranda}
{\rm Integrand}^{(2)}_{{\rm dcut-A}}
=
\frac{
- \ft12 S_{123} \tr_4 [\bar{P}_6 P_1 \bar{K}_2 K_1 \bar{P}_3 P_4 \bar{K}_4 K_3]
}{
K_1^2 K_2^2 K_3^2 K_4^2 S_{3 K_1} S_{1 K_2} S_{K_2 K_3}S_{4 K_4} S_{6 K_3}
}
\, ,
\end{align}
with the assumed momentum conservation conditions $K_1 + K_2 + P_1 + P_2 + P_3 = 0$, $K_3 + K_4 + P_4 + P_5 + P_6$, and 
$K_1 + K_2 + K_3 + K_4 = 0$. Calculating the trace with {\tt FeynCalc} and expressing Lorentz products in terms of the  propagators 
occurring in \re{2LoopIntegranda} as well as a set of irreducible scalar products $\{ S_{4,-K_1}, S_{6,-K_2}, S_{1,-K_3}, S_{3,-K_4} \}$, 
we find
\begin{align}
- 2 
&
\times \tr_4 [\bar{P}_6 P_1 \bar{K}_2 K_1 \bar{P}_3 P_4 \bar{K}_4 K_3]
\nonumber\\
&
=
[S_{34}S_{61}S_{123} + S_{12}S_{45}S_{234} + S_{23}S_{56}S_{612} - S_{123} S_{234} S_{612}] S_{K_2K_3}
\nonumber\\
&
-
[S_{12} S_{45} - S_{123} S_{612}] [ S_{1,-K_3} S_{4,-K_1} + S_{1K_2} S_{4K_4}]
\nonumber\\
&
+
[S_{12} S_{45} - S_{123} S_{612}] [S_{3,-K_4} S_{6,-K_2} + S_{3K_1} S_{6, K_3}]
\nonumber\\
&
+
S_{45} S_{56} [ S_{1,-K_3} S_{3K_1} + S_{1K_2} S_{3,-K_4} ]
\nonumber\\
&
+
S_{12} S_{23} [ S_{4K_4} S_{6,-K_2} + S_{4,-K_1} S_{6K_3} ]
\nonumber\\
&
+
S_{61} S_{123} [ S_{3K_1} S_{4K_4} - S_{3,-K_4} S_{4,-K_1} ]
\nonumber\\
&
+
S_{34} S_{123} [ S_{1K_2} S_{6K_3} - S_{1,-K_3} S_{6,-K_2} ]
\, .
\end{align}
Matching the propagator and numerator structures to the set of basis integrals in Fig.\ \ref{TwoLoopIntergalsPic} immediately 
provides the form of the accompanying coefficients for integrals $\mathcal{I}^{(2)}_\alpha$ with indices $\alpha = 1,3,4,9,12,13$. 
Some of them enter more than once in various permutations. They are summarized in Eq.\ \re{2LoopCoefficients} below.

%%%%%%%%%%%%%%%%%%%%%%%%%%%%%%%%%%%%%%%%%%%%%%%%%%%%%%%%%%%%%%%%%%%%%
%            Figure
%%%%%%%%%%%%%%%%%%%%%%%%%%%%%%%%%%%%%%%%%%%%%%%%%%%%%%%%%%%%%%%%%%%%%
%%%%%%%%%%%%%%%%%%%%%%%%%%%%%%%%%%%%%%%%%%%%%%%%%%%%%%%%%%%%%%%%%%%%%

\subsubsection{Graph B}

Moving on to the double cut B in Fig.\ \ref{TwoLoopPic}, it reads
\begin{align}
\label{2loopCutb}
\big. \mathcal{A}_6^{(2)} \big|_{{\rm dcut-B}}
&
=
\delta^{(4)} \left( Q_{123456} \right)
\delta^{(4)} \left( \bar{Q}_{123456} \right)
\int \prod_{i=1}^4 d^2 \eta_{K_i} d^2 \bar\eta_{K_i}
\\
&\times
\delta^{(4)} \left( Q_{6,-K_1,-K_2,-K_3,-K_4} \right)
\!
\delta^{(4)} \left( \bar{Q}_{6,-K_1,-K_2,-K_3,-K_4} \right)
\!
\delta^{(4)} \left( Q_{345 K_3 K_4} \right)
\!
\delta^{(4)} \left( \bar{Q}_{345 K_3 K_4} \right)
\nonumber\\[2mm]
&
\times
\widehat{\mathcal{A}}_4^{(0)} (P_1, P_2, K_1, K_2)
\widehat{\mathcal{A}}_5^{(0)} (-K_1, -K_2, P_6,-K_3, - K_4)
\widehat{\mathcal{A}}_5^{(0)} (K_3, K_4, P_3, P_4, P_5)
\, . \nonumber
\end{align}
Its projection on the component in question is determined by the equation
\begin{align}
\VEV{\VEV{
\big. \mathcal{A}_6^{(2)} \big|^{\rm sing}_{{\rm dcut-B}}
}}
&
=
\VEV{\VEV{
\frac{S_{12}}{S_{2 K_1}}
\frac{\bra{Q_6} \bar{K}_2 K_1 \bar{K}_4 K_3 | \bar{Q}_6]}{S_{6,-K_2} S_{K_1 K_4} S_{6,-K_3}}
\frac{\bra{Q_3} \bar{K}_4 K_3 \bar{P}_5 P_4 | \bar{Q}_3]}{S_{3 K_4} S_{34} S_{5 K_3}}
}}
\, ,
\end{align}
such that after the use of Eq.\ \re{IntQQ2P}, the two-loop integrand detected by this discontinuity becomes
\begin{align}
\label{2LoopIntegrandb}
{\rm Integrand}^{(2)}_{{\rm dcut-B}}
=
\frac{
\ft12 \tr_4 [\bar{P}_6 P_1 \bar{P}_2 P_3 \bar{K}_4 K_3 \bar{P}_5 P_6 \bar{K}_2 K_1 \bar{K}_4 K_3]
}{
K_1^2 K_2^2 K_3^2 K_4^2 S_{2 K_1} S_{6, - K_2} S_{6, - K_3} S_{K_1 K_4}S_{3 K_4} S_{5 K_3}
}
\, .
\end{align}
Here, as before, the momentum conservation conditions are implied, $K_1 + K_2 + P_1 + P_2 = 0$, $K_1 + K_2 + K_3 + K_4 - P_6 = 0$, 
and $K_3 + K_4 + P_3 + P_4 + P_5 = 0$. The trace is easily evaluated, but it needs some `combing' to cast in a pleasant-looking form.
We find, after some manipulations,
\begin{align}
2 &
\times \tr_4 
[
\bar{P}_6 P_1 \bar{P}_2 P_3 \bar{K}_4 K_3 \bar{P}_5 P_6 \bar{K}_2 K_1 \bar{K}_4 K_3]
\nonumber\\
&
=
-
S_{612} S_{6,-K_3}  S_{K_1 K_4}
[ S_{234} S_{612} S_{123}-S_{61} S_{34} S_{123}-S_{12} S_{45} S_{234}-S_{23} S_{56} S_{612} ]
\nonumber\\
&
-
[ S_{12} S_{45} - S_{123} S_{612}] 
[
S_{612} S_{2K_1} S_{5K_3} S_{6,-K_3} 
+
S_{612} S_{56,-K_2} S_{6,-K_3}  S_{2,-K_4} 
+
2
S_{12}  S_{6,-K_2} S_{5K_3}  S_{2,-K_4}
]
\nonumber\\
&
-
S_{12} S_{6,-K_2} S_{6,-K_3} S_{3 K_4} [2 S_{61} S_{34} - S_{234} S_{612}]
\nonumber\\
&
+
S_{34}  S_{6,-K_2} S_{6,-K_3}  S_{2,-K_4} [2 S_{12} S_{45}-S_{123} S_{612} ]
\nonumber\\
&
-
S_{612} S_{3, -K_1} S_{6,-K_3}^2 [ S_{61} S_{34}-S_{234} S_{612} ]
\nonumber\\
&
-
S_{56} S_{612}^2 
[
S_{3, -K_1} S_{6,-K_3}  S_{2,-K_4}
-
S_{2K_1} S_{6,-K_3} S_{3 K_4}
]
\nonumber\\
&
+
S_{61} S_{12} S_{612} 
[
S_{56,-K_2} S_{6,-K_3} S_{3 K_4}
+
S_{3, -K_1} S_{5K_3} S_{6,-K_3}
]
\nonumber\\
&
+
2 S_{61} S_{12}^2 S_{6,-K_2} S_{5K_3} S_{3 K_4}
-
S_{23} S_{12} S_{612}  S_{6,-K_2} S_{5K_3} S_{6,-K_3}
\nonumber\\
& 
-
S_{23} S_{612}^2 S_{56,-K_2} S_{6,-K_3}^2
+
S_{23} S_{34} S_{612}  S_{6,-K_2} S_{6,-K_3}^2
+
S_{34} S_{45} S_{612} S_{2K_1} S_{6,-K_3}^2
\, .
\end{align}
Merging this with the cut of the previous section, we confirm the values of the coefficients for $\alpha = 1,3,4,9,12,13$, and determine
a few new ones with $\alpha = 2,5,6,10,11$.

\subsubsection{Graph C}

Next, we consider the double cut C. It is
\begin{align}
\label{2loopCutc}
\big. \mathcal{A}_6^{(2)} \big|_{{\rm dcut-C}}
&
=
\delta^{(4)} \left( Q_{123456} \right)
\delta^{(4)} \left( \bar{Q}_{123456} \right)
\int \prod_{i=1}^4 d^2 \eta_{K_i} d^2 \bar\eta_{K_i}
\\
&\hspace{-1cm}\times
\delta^{(4)} \left( Q_{6,-K_1,-K_2,-K_3,-K_4} \right)
\delta^{(4)} \left( \bar{Q}_{6,-K_1,-K_2,-K_3,-K_4} \right)
\delta^{(4)} \left( Q_{3456 K_3 K_4} \right)
\delta^{(4)} \left( \bar{Q}_{3456 K_3 K_4} \right)
\nonumber\\[2mm]
&
\times
\widehat{\mathcal{A}}_4^{(0)} (P_1, P_2, K_1, K_2)
\widehat{\mathcal{A}}_4^{(0)} (-K_1, - K_2,-K_3, - K_4)
\widehat{\mathcal{A}}_6^{(0)} (K_3, K_4, P_3, P_4, P_5, P_6)
\, . \nonumber
\end{align}
It contains the six-leg amplitude. For this discontinuity, since it appears as a right-most component, it suffices to use its simplified 
form \re{LessSymA6}. The projection
\begin{align}
\VEV{\VEV{
\big. \mathcal{A}_6^{(2)} \big|^{\rm sing}_{{\rm dcut-C}}
}}
&
=
\VEV{\VEV{
\frac{S^2_{12}}{S_{2 K_1} S_{K_1 K_4}}
\frac{\bra{Q_3} \bar{K}_4 K_3 | \bar{Q}_6] \bra{Q_6} \bar{P}_5 P_4 | \bar{Q}_3]}{S_{3 K_4} S_{34} S_{56} S_{6 K_3}}
}}
\, ,
\end{align}
is easily calculated with the result for the integrand
\begin{align}
\label{2LoopIntegrandc}
{\rm Integrand}^{(2)}_{{\rm dcut-C}}
=
\frac{
\ft12 S_{12} \tr_4 [\bar{P}_6 K_3 \bar{K}_4 P_3 \bar{P}_2 P_1]
}{
K_1^2 K_2^2 K_3^2 K_4^2 S_{2 K_1} S_{K_1 K_4} S_{3 K_4} S_{6 K_3}
}
\, ,
\end{align}
with implied momentum conservation $K_3 + K_4 = P_1 + P_2$. The trace gives
\begin{align}
\tr_4 [\bar{P}_6 K_3 \bar{K}_4 P_3 \bar{P}_2 P_1]
=
[S_{12} S_{34} - S_{612} S_{123}] S_{1,-K_3}
+
S_{61} S_{12} S_{3 K_4}
+
S_{12} S_{23} S_{6 K_3}
\, ,
\end{align}
and confirm the previously found coefficients with labels $\alpha = 2,8$.

\subsubsection{Graph D}

Finally, we arrive at the cut D. It detects the integrals with $\alpha = 1,3,6,7,8,9,10,11,12,13$. Notice, however, all of their
coefficients are completely fixed by the discontinuities in A, B, and C, except for $c^{(2)}_7$. As we will demonstrate in the 
next section, this cut is, in fact, not needed explicitly. We may merely rely on its very generic property of not possessing a 
multiparticle factorization channel with the loop-momentum-dependent $S_{3, - K_1, - K_4}$-denominator by virtue of the 
same property of the middle, MHV-like six-particle amplitude. Nevertheless, for completeness, we present the corresponding 
calculation in order to (i) demonstrate subtleties of the six-dimensional calculation and (ii) exhibit similarities/differences with 
the four-dimensional case reported in the Appendix \ref{Appendix4Dcuts}.

In the double cut D, 
\begin{align}
\label{2loopCutd}
\big. \mathcal{A}_6^{(2)} \big|_{{\rm dcut-D}}
&
=
\delta^{(4)} \left( Q_{123456} \right)
\delta^{(4)} \left( \bar{Q}_{123456} \right)
\int \prod_{i=1}^4 d^2 \eta_{K_i} d^2 \bar\eta_{K_i}
\\
&\hspace{-1cm}\times
\delta^{(4)} \left( Q_{6,-K_1,-K_2,-K_3,-K_4} \right)
\delta^{(4)} \left( \bar{Q}_{6,-K_1,-K_2,-K_3,-K_4} \right)
\delta^{(4)} \left( Q_{3456 K_3 K_4} \right)
\delta^{(4)} \left( \bar{Q}_{3456 K_3 K_4} \right)
\nonumber\\[2mm]
&
\times
\widehat{\mathcal{A}}_4^{(0)} (P_1, P_2, K_1, K_2)
\widehat{\mathcal{A}}_6^{(0)} (-K_1, - K_2, P_3, -K_3, - K_4, P_6)
\widehat{\mathcal{A}}_4^{(0)} (K_3, K_4, P_4, P_5)
\, , \nonumber
\end{align}
the six-leg amplitude appears sandwiched between the four-leg ones; thus, we cannot rely on its simplified expression \re{LessSymA6}, 
instead, we have to use its full form \re{treeA6}. But then there is a catch, since the projector $\Pi_{36}$ does not provide a clean extraction 
of the full structure in question. Namely, the cc-portion of the tree amplitude yields an additive trace squared
\begin{align}
\label{SymGraphD}
\VEV{\VEV{
\mathcal{W}_{12} \widebar{\mathcal{W}}_{45} + \mathcal{W}_{45} \widebar{\mathcal{W}}_{12}
}}
=
S_{12} S_{23} S_{34} S_{45} S_{56}
+
\big[\!
\tr_4 [\bar{P}_1 P_2 \bar{P}_3 P_4 \bar{P}_5 P_6] 
\big]^2
\, .
\end{align}
The latter is a polynomial in the cross-ratios \re{CrossRatios}, $\tr_4 [\bar{P}_1 P_2 \bar{P}_3 P_4 \bar{P}_5 P_6] = 
S_{123} S_{123} S_{234} (1 - U_1 - U_2 - U_3)$. This complicates the identification of the true integrand corresponding 
to the component in question. We need to get rid of this extra term. It cannot be done with a covariant six-dimensional projector 
due to the lack of a rank-two tensor that can lower/raise ${\rm SU}^\ast(4)$ Lorentz indices. One option is to pass to the 
two-dimensional component form. However, we would like to stay in six dimensions and, nevertheless, alleviate this 
predicament. This can be done by subtracting the unwanted term by hand. Namely, the double trace structure of the second 
term in Eq.\ \re{SymGraphD} tells us that inside the unitarity cut, one of the two factors will preserve its form and depend 
only on the external momenta, while the other one will possess quartic dependence on momenta of the cut lines `sprinkled' 
with the external gluon momenta $P_{3,6}$. Then it is not hard to see that the combination of traces determining the numerator
of the graph D is equivalent to the single trace
\begin{align}
\ft12 \tr_4 [P_3 \bar{P}_2 P_1 \bar{P}_6 K_2 \bar{K}_1 P_3 \bar{P}_4 P_5 \bar{P}_6 K_3 \bar{K}_4]
&=
\ft12\tr_4[\bar{P}_6 K_2 \bar{K}_1 P_3 \bar{P}_4 P_5] \ft12 \tr_4[P_6 \bar{K}_3 K_4 \bar{P}_3 P_2 \bar{P}_1]
\nonumber\\
&
+
\ft12 \tr_4[P_6 \bar{K}_2 K_1 \bar{P}_3 P_2 \bar{P}_1] \ft12\tr_4[\bar{P}_6 K_3 \bar{K}_4 P_3 \bar{P}_4 P_5]
\nonumber\\
&
-
\ft12 \tr_4[P_6 \bar{P}_1 P_2 \bar{P}_3 P_2 \bar{P}_1] \ft12 \tr_4[\bar{P}_6 K_2 \bar{K}_1 P_3 \bar{K}_4 K_3]
\, . \nonumber
\end{align}
This is exactly the Dirac string that one encounters in the four-dimensional calculation of Appendix \ref{Appendix4Dcuts},
except that every momentum is uplifted from four to six dimensions! The calculation of this trace yields a very lengthy
expression
\begin{align}
2 
&
\times \tr_4 [P_3 \bar{P}_2 P_1 \bar{P}_6 K_2 \bar{K}_1 P_3 \bar{P}_4 P_5 \bar{P}_6 K_3 \bar{K}_4]
\\
&
=
S_{34} S_{61} [S_{12} S_{45}-S_{123} S_{612}]^2
\nonumber\\
&
-
S_{3,-K_{1},-K_{4}} [S_{12} S_{45}-S_{123} S_{612}] [S_{34} S_{61} S_{123}+S_{12} S_{45} S_{234} + S_{23} S_{56} S_{612} - S_{612} S_{123} S_{234}] 
\nonumber\\
&
+
S_{6,-K_{3}} S_{34} [S_{12} S_{45}-S_{123} S_{612}] [S_{12} S_{45}+(S_{23}-S_{123}) S_{612}]
\nonumber\\
&
+
S_{3,-K_{1}} S_{61} [S_{12} S_{45}-S_{123} S_{612}] [S_{12} S_{45}+(S_{56}-S_{123}) S_{612}]
\nonumber\\
&
+
[S_{12} S_{45}-S_{123} S_{612}]^2 [S_{34} S_{1,-K_{3}} +  S_{61}S_{4,-K_{1}}]
\nonumber\\
&
+
[S_{12} S_{45}-S_{123} S_{612}]^2[S_{1,-K_{3}} S_{4,-K_{1}}  + S_{2K_{1}} S_{4K_{4}} ]
\nonumber\\
&
+
S_{34}  S_{61} S_{123} [S_{12} S_{45}-S_{123} S_{612}] [S_{3,-K_{4}} + S_{6,-K_{2}}]
\nonumber\\
&
+
S_{123} [S_{12} S_{45}-S_{123} S_{612}][S_{61} S_{3,-K_{4}} S_{4,-K_{1}} + S_{34} S_{1,-K_{3}} S_{6,-K_{2}} ]
\nonumber\\
&
+
S_{3,-K_{4}} S_{6,-K_{2}} [S_{12} S_{45}-S_{123} S_{612}] [S_{23} S_{56}-S_{123} S_{234}] 
\nonumber\\
&
+
S_{3,-K_{1}} S_{6,-K_{3}} [S_{12} S_{45}-S_{123} S_{612}] [S_{12} S_{45}+S_{34} S_{61}+(S_{23}+S_{56}-S_{123}-S_{234}) S_{612}]
\nonumber\\
&
+
S_{4,-K_{1}} S_{6,-K_{3}} [S_{12} S_{45}-S_{123}S_{612}] [S_{12} S_{45}+(S_{23}-S_{123}) S_{612}] 
\nonumber\\
&
+
S_{1,-K_{3}} S_{3,-K_{1}} [S_{12} S_{45}-S_{123} S_{612}] [S_{12} S_{45}+(S_{56}-S_{123}) S_{612}] 
\nonumber\\
&
+
S_{6,-K_{2}} S_{6,-K_{3}} S_{34} [S_{12} S_{45} (S_{23}+S_{123})+(S_{23}-S_{123}) S_{123} S_{612}]
\nonumber\\
&
+
S_{3,-K_{1}} S_{3,-K_{4}} S_{61} [S_{12} S_{45} (S_{56}+S_{123})+(S_{56}-S_{123}) S_{123} S_{612}]
\nonumber\\
&
-
[S_{12} S_{45}-S_{123} S_{612}]
[
S_{12} S_{23}  S_{4K_{4}} S_{6,-K_{2}}
+
S_{12} S_{61} S_{3,-K_{1}} S_{4K_{4}} 
\nonumber\\
&\hspace{70.5mm}
+
S_{34} S_{45} S_{2K_{1}} S_{6,-K_{3}}
+
S_{45} S_{56} S_{2K_{1}} S_{3,-K_{4}}
]
\, . \nonumber
\end{align}
Here, contributions linear in Levi-Civita symbols stemming from individual traces \re{6DtraceLC} do not show up as we explicitly 
verified\footnote{This had to be done by hand since relying on {\tt FeynCalc} would not help as the latter operates with 
four-dimensional Dirac algebra or its $\varepsilon$-vicinity only. While the parity-even parts are the same in any dimension 
(except for the overall trace normalization), the parity-odd parts differ.} that their product vanishes due to the linear dependence 
of momenta involved. 

The contribution to the integrand from this cut reads
\begin{align}
\label{2LoopIntegrandd}
{\rm Integrand}^{(2)}_{{\rm dcut-D}}
=
\frac{
\ft12 \tr_4 [P_3 \bar{P}_2 P_1 \bar{P}_6 K_2 \bar{K}_1 P_3 \bar{P}_4 P_5 \bar{P}_6 K_3 \bar{K}_4]
}{
K_1^2 K_2^2 K_3^2 K_4^2 S_{2K_1} S_{3, -K_1} S_{3,-K_4} S_{6,-K_3} S_{6, -K_2} S_{5K_3} S_{4K_4}
}
\, .
\end{align}
However, contrary to other graphs, it does not immediately yield the coefficients of the chosen integral basis because of the `missing' 
denominator $S_{3, - K_1, - K_4}$. As we mentioned at the beginning of this section, the MHV-like nature of the middle six-particle 
amplitude prevents it from showing up in this cut because the former does not possess this three-particle factorization channel. 
This implies that without a dimension-dependent reduction procedure, potentially inducing surface terms for formally irreducible 
integrands, as well as Gram-determinant relations, as explained at length \cite{Kosower:2011ty}, one would have to resort to numerical 
arguments\footnote{We are indebted to Lance Dixon for the explanation of the calculation in the four-dimensional case, and sharing 
an unpublished Maple code of the corresponding test.} to confirm the ansatz for the cut in terms of scalar integrals 
in Fig.\ \ref{TwoLoopIntergalsPic}. We would like to avoid this, be a bit more methodological, and perform this extraction analytically. 
To this end, we turn to the leading singularity analysis of this topology, i.e., additionally cutting the graph D in Fig.\ \ref{TwoLoopPic} 
horizontally across all $t$-channel propagators. An attentive reader would realize that this cut vanishes identically. However, this 
information alone suffices. It implies that the three-particle $S_{3, - K_1, - K_4}$-pole in the integrand of the two-loop amplitude, 
which is present in individual basis integrals, is in fact spurious. This gives direct access to the so far unconstrained coefficient of $\mathcal{I}_7$. 

\subsubsection{Heptacut}

As we just said, to extract the coefficient $c^{(2)}_7$ efficiently, we need to freeze as many degrees of freedom as 
possible. A look at the corresponding graph in Fig.\ \ref{TwoLoopIntergalsPic} suggests that, as we just mentioned 
at the end of the previous section, we need to cut all seven propagators. So we turn to the analysis of the heptacut 
\cite{Kosower:2010yk,Kosower:2011ty,Mastrolia:2011pr,Larsen:2012sx,Badger:2012dp} within the context of the 
leading-singularity approach of Refs.\ \cite{Buchbinder:2005wp,Cachazo:2008vp,Cachazo:2008hp}. This consideration 
is analogous to the one for the four-dimensional setup that we perform for the reader's convenience in Appendix \ref{Appendix4Dcuts}.

In this section, we will adopt a somewhat mixed notation for the six-dimensional vectors $P_i = (p_i^{\dot\alpha\alpha}, y_{i,i+1})$: 
its four-dimensional components are given in terms of four-dimensional Weyl spinors, while the out-of-four-dimensional remainder
is written via the two-dimensional dual variables $y$. The squared norm of the six-vector is then
\begin{align}
P_i^2 \equiv \ft12 \tr_2 [p_i p_i] - y_{i,i+1}^2
\, .
\end{align}

\begin{figure}[t]
\begin{center}
\mbox{
\begin{picture}(0,140)(120,0)
\put(0,0){\insertfig{8}{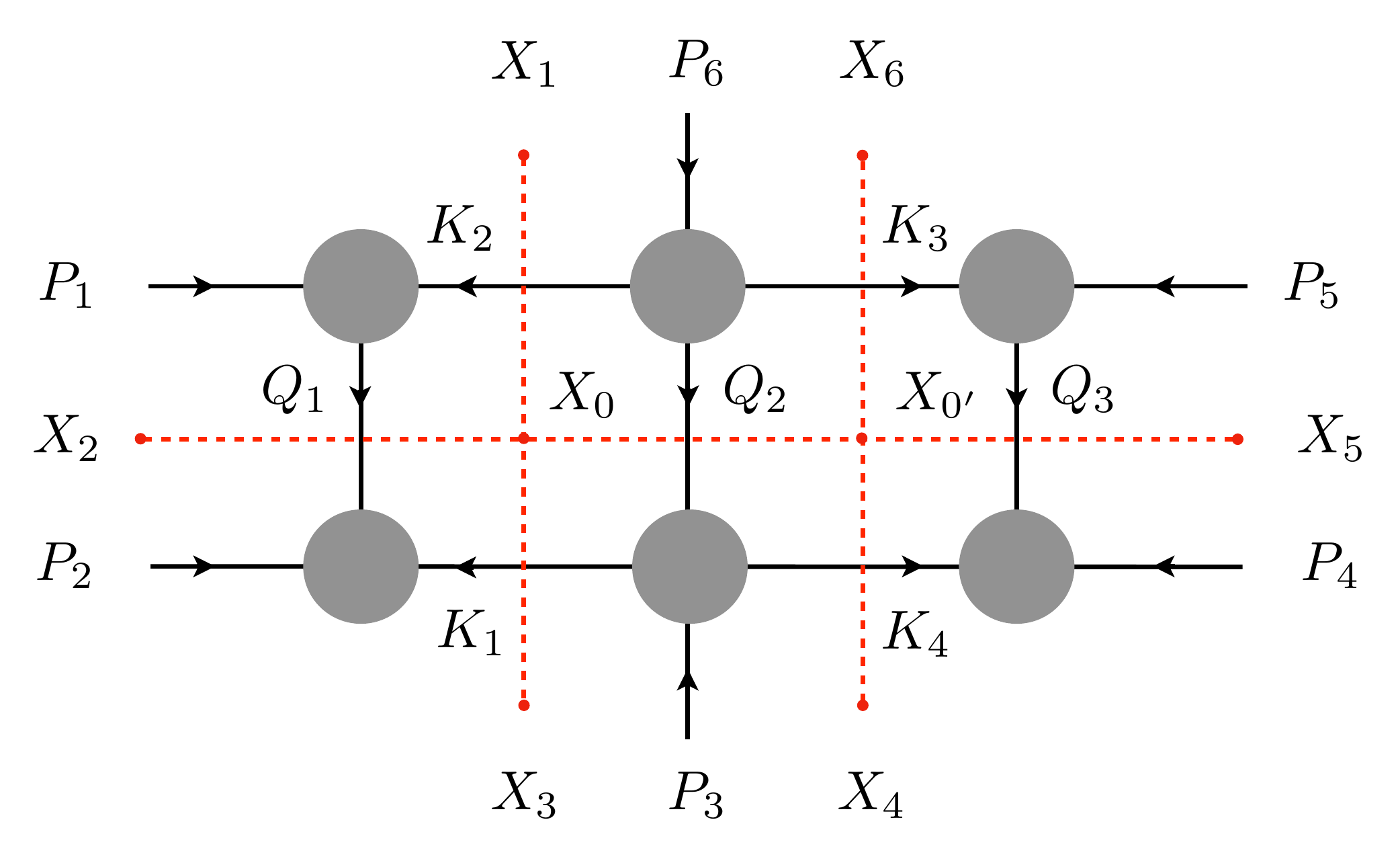}}
\end{picture}
}
\end{center}
\caption{\label{HeptacutPic} The heptacut of the two-loop integrals determining the coefficient of the integral $\mathcal{I}^{(2)}_7$. In 
dashed lines, we exhibit the dual graph.}
\end{figure}

Formally, the heptacut, shown in Fig.\ \ref{HeptacutPic}, fixes 7 out of $2 \times 6$ loop integrations, leaving one degree of freedom 
residing in four dimensions and $2+2$ extra-dimensional coordinates $y_{0}$ and $y_{0^\prime}$ being unfixed. Notice, however, that 
the latter are eventually set to zero upon the dimensional reduction \re{DRmeasure}, i.e., they are not dynamical. Then, by generalizing 
the massless case worked out in Appendix \ref{Appendix4Dheptacut}, the solutions to the mass-shell $Q_{1,3}^2 = 0$ is parametrized 
by the variables $z_{L,R}$ as well as the $y_{0}$- and $y_{0^\prime}$-coordinates 
\begin{align}
\label{Q1}
Q_1 
&
= 
\left( z_L \ket{1} [2| + \frac{y_{02}^2}{z_L \vev{12} [12]} \ket{2} [1|, y_{02} \right)
\, , \\
\label{Q3}
Q_3 
&
= 
\left( z_R \ket{4} [5| + \frac{y_{0^\prime 5}^2}{z_R \vev{45} [45]} \ket{5} [4|, y_{5 0^\prime} \right)
\, .
\end{align}
We will use different auxiliary vectors to parametrize mass effects in external legs, which are natural choices for the corresponding
individual $P_{1,2}$ and $P_{4,5}$ pairs
\begin{align}
\label{P12}
P_1 
&
= \left( \ket{1} [1| - \frac{y_{12}^2}{\vev{12} [12]} \ket{2} [2|, y_{12} \right)
\, , \quad
P_2 
= \left( \ket{2} [2| - \frac{y_{23}^2}{\vev{12} [12]} \ket{1} [1|, y_{23} \right)
\, , \\
\label{P45}
P_4
&
= \left( \ket{4} [4| - \frac{y_{45}^2}{\vev{45} [45]} \ket{5} [5|, y_{45} \right)
\, , \quad
P_5
= \left( \ket{5} [5| - \frac{y_{56}^2}{\vev{45} [45]} \ket{4} [4|, y_{56} \right)
\, .
\end{align}
Then, the cut conditions $K_{1,2,3,4}^2 = 0 $ on legs $K_{1,2,3,4}$ yield two equations each for the two-vectors 
$y_0$ and $y_{0^\prime}$,
\begin{alignat}{3}
&
y_{12} \cdot y_{02} = 0 \, , \quad 
&&
y_{23} \cdot y_{02} = 0 
&& 
\quad \to \quad y_0 = y_2
\, , \\
&
y_{45} \cdot y_{0^\prime 5} = 0 \, , \quad 
&&
y_{56} \cdot y_{0^\prime 5} = 0 
&& 
\quad \to \quad y_{0^\prime} = y_5
\, .
\end{alignat}
At first sight, this might appear as though we overconstrained the system. While indeed these might not be the most general solutions, 
and their simplicity partially reflects the choice of reference spinors in our parametrization in Eqs.\ \re{P12} and \re{P45},
they nevertheless suffice our needs since internal integrations are eventually reduced down to four space-time dimensions by the 
compactification \re{DRmeasure}. So they are consistent with Eq.\ \re{6Dto4D}. On these solutions, the null six-vectors $Q_{1,3}$ are 
identical to their four-dimensional counterparts \re{4Dq13solutions}. Like in four dimensions, these $t$-channel momenta vanish for 
$z_{L,R} = 0$, respectively. These are locations for the `composite' leading singularities \cite{Buchbinder:2005wp,Cachazo:2008vp}, which are used for 
integration contours in the unconstrained degree of freedom $z_L$ or $z_R$.

This observation has profound consequences for the extraction of the $c^{(2)}_7$ coefficient. Namely, the three-particle Madelstam
invariants vanish with $z_{L, R}$,
\begin{align}
S_{23, -K_4} = 2 Q_1 \cdot Q_2 \sim z_L
\, , \qquad
S_{34, -K_1} = 2 Q_3 \cdot Q_2 \sim z_R
\, .
\end{align}
The final degree of freedom $z_L$ or $z_R$ is integrated over a contour in the complex plane. Integration over a small circle
around the `composite' singularities $z_L = 0$ or $z_R = 0$ yields identical equations for the expansion 
coefficient in question. So effectively, we compute an octacut \cite{Buchbinder:2005wp}. Let us focus on $z_L = 0$. In this case, we 
have $K_2 = - P_1$ and $K_1 = - P_2$, so that the loop momentum-dependent invariants become
\begin{align}
S_{3,-K_1} = S_{23}
\, , \qquad
S_{6,-K_2} = S_{61}
\, , \qquad
S_{34,- K_1} = S_{234}
\, , \qquad
S_{23, -K_4} = 0
\, .
\end{align}
As we explained in Appendix \ref{Appendix4Dheptacut}, this immediately yields
\begin{align}
c^{(2)}_7 = - \sum_{
\genfrac{.}{.}{0pt}{2}{(543216)}{(456123)}
} \left[ c^{(2)}_{10} + c^{(2)}_{11} \right]
\, ,
\end{align}
and thus concludes our analysis of all required cuts. The basis of integrals and their coefficients are now completely fixed.

%%%%%%%%%%%%%%%%%%%%%%%%%%%%%%%%%%%%%%%%%%%%%%%%%%%%%%%%%%%%%%%%%%%%%
%            Figure
%%%%%%%%%%%%%%%%%%%%%%%%%%%%%%%%%%%%%%%%%%%%%%%%%%%%%%%%%%%%%%%%%%%%%
%%%%%%%%%%%%%%%%%%%%%%%%%%%%%%%%%%%%%%%%%%%%%%%%%%%%%%%%%%%%%%%%%%%%%

\subsection{Integrals and coefficients}
\label{IntBasisSection}

Summarizing our findings from the previous sections, the assembly equation for the two-loop amplitude is 
\begin{align}
\label{2loopA6}
\mathcal{A}^{(2)}_{6} 
=
\ft14 \mathcal{A}^{(0)}_{6}  \sum_{\sigma_6 \cup \bar\sigma_6} 
\Big[
&
\ft14 c^{(2)}_1 \mathcal{I}_1^{(2)}
+ c^{(2)}_2 \mathcal{I}_2^{(2)}
+ \ft12 c^{(2)}_3 \mathcal{I}_3^{(2)}
+ \ft12 c^{(2)}_4 \mathcal{I}_4^{(2)}
+ c^{(2)}_5 \mathcal{I}_5^{(2)}
+ c^{(2)}_6 \mathcal{I}_6^{(2)}
\\
+ 
&
\ft14 c^{(2)}_7 \mathcal{I}_7^{(2)}
+ \ft12 c^{(2)}_8 \mathcal{I}_8^{(2)}
+ c^{(2)}_9 \mathcal{I}_9^{(2)}
+ c^{(2)}_{10} \mathcal{I}_{10}^{(2)}
+ c^{(2)}_{11} \mathcal{I}_{11}^{(2)}
+ \ft12 c^{(2)}_{12} \mathcal{I}_{12}^{(2)}
+ \ft12 c^{(2)}_{13} \mathcal{I}_{13}^{(2)}
\Big] 
, \nonumber
\end{align}
where the sum runs over the six cyclic permutations \re{CyclicPerms} and their reflections 
\begin{align}
\label{CyclicPermsReflection}
\bar\sigma_6 = \{ (654321),(543216),(432165),(321654),(216543),(165432)\}
\, .
\end{align}
They form the dihedral group D$_6$. The coefficients are in turn
\begin{align}
\label{2LoopCoefficients}
c^{(2)}_1 &= 
S_{12} S_{45} S_{123} S_{234}  
+ 
S_{23} S_{56} S_{123} S_{612}
+
S_{123}^2 [ S_{34} S_{61} - S_{234} S_{612} ]
\, , \nonumber\\
c^{(2)}_2 &= 
2 S_{23} S_{12}^2 
\, , \nonumber\\
c^{(2)}_3 &= 
S_{123} [ S_{123} S_{234} - S_{23} S_{56} ] 
\, , \nonumber\\
c^{(2)}_4 &= 
S_{61} S_{123}^2 
\, , \nonumber\\
c^{(2)}_5 &= 
S_{12} [ S_{123} S_{234} - 2 S_{23} S_{56} ] 
\, , \nonumber\\
c^{(2)}_6 &= 
- S_{61} S_{56} S_{123} 
\, , \nonumber\\
c^{(2)}_7 &= 
2 S_{123} S_{234} S_{612} 
- 
4 S_{12} S_{45} S_{234} 
- 
S_{23} S_{56} S_{612} 
- 
S_{34} S_{61} S_{123}
\, , \nonumber\\
c^{(2)}_8 &= 
2 S_{12} [ S_{123} S_{612} - S_{12} S_{45} ] 
\, , \nonumber\\
c^{(2)}_9 &= 
S_{45} S_{56} S_{123} 
\, , \nonumber\\
c^{(2)}_{10} &= 
S_{56} [ 2 S_{12} S_{45} - S_{123} S_{612} ] 
\, , \nonumber\\
c^{(2)}_{11} &= 
S_{23} S_{34} S_{612} 
\, , \nonumber\\
c^{(2)}_{12} &= 
S_{123} [ S_{123} S_{612} - S_{12} S_{45} ] 
\, , \nonumber\\
c^{(2)}_{13} &= - S_{123}^2 S_{34}
\, .
\end{align}
As it was obvious in the course of the calculation, the above results are an uplift of four-dimensional invariants to six
dimensions, i.e., $s_{ij\dots} \to S_{ij\dots}$. This list, together with the integrals in Fig.\ \ref{TwoLoopIntergalsPic}, is the 
main result of this work. 

\section{Conclusion}

In this paper, we constructed a basis of integrals and their accompanying coefficients, which they enter with into the off-shell 
six-leg amplitude to the two-loop order. In the on-shell limit, it corresponds to the MHV six-gluon amplitude. We found that all of 
its ingredients can be obtained by an uplift of the four-dimensional Lorentz invariants to six. The limitations of the current method 
did not allow us to extract the parity-odd part of the amplitude since we favored covariance over the component formulation of the
$\mathcal{N} = (1,1)$ sYM. However, in Ref.\ \cite{Cachazo:2008hp} it was demonstrated that the parity-odd part drops out from 
the logarithm of the amplitude at the conformal point of the theory. We have no reason to believe that this will not be the case on 
the Coulomb branch of the theory as well. We were also not sensitive to the potential $\mu$-terms; however, these are of no 
relevance for the problem at hand since the loop integrations are performed strictly in four dimensions after the generalized 
dimensional reduction: all infrared divergences are regulated by the nonvanishing off-shellness.

The next task at hand is to evaluate contributing scalar Feynman integrals analytically. Until very recently, this would seem like
an impossible task. The conventional method of differential equations \cite{Kotikov:1990kg,Gehrmann:1999as,Henn:2013pwa}, 
being the most efficient technique to date for high-loop integrations, is not well-suited for the problem at hand, since the off-shellness 
makes integral reductions swell beyond control, even with the modular arithmetic approach \cite{Belitsky:2024jhe}. There exists a 
way to make the formalism more efficient by using the dual conformal invariance property of off-shell Feynman integrals. Such 
a modification was proposed in Ref.\ \cite{Caron-Huot:2014lda}. This deserves an independent exploration.

Recently, we offered a complementary formalism \cite{Belitsky:2025sin} to analyze the near mass-shell limit of contributing integrals. 
This was accomplished by building on the technique introduced in Ref.\ \cite{Salvatori:2024nva} to iteratively calculate finite Feynman 
integrals by `factoring out' infrared singularities first. However, in its original form, the formalism did not account for the fact that loop 
integrals are dual-conformally invariant, and subsequent simplifications required extensive use of the motivic theory of recursive integrals
\cite{Goncharov:2009lql,Goncharov:2010jf}. Presently, we can do better since we can rely on the dual conformally-invariant formulation 
\cite{Bork:2025ztu} of the Method of Regions \cite{Beneke:1997zp}, which provides individual regions in terms of dual conformal integrals. 
This significantly reduces the required manipulations on the way to a fully analytical prediction of the six-leg amplitude of nearly massless 
W-bosons. These results will be reported elsewhere.

\begin{acknowledgments}
We would like to thank Lance Dixon, Radu Roiban, and Cristian Vergu for insightful correspondence. We are particularly indebted
to Lance Dixon for sharing his unpublished code for the six-gluon MHV amplitude.
\end{acknowledgments}

\appendix

\section{Six-dimensional Dirac matrices}
\label{6DAppendix}

Our conventions for the six-dimensional Dirac algebra are borrowed from \cite{Belitsky:2024rwv}, where the reader can find all the 
necessary details. Presently, we only need to supplement them with a list of traces used in the present consideration. The Dirac 
matrices
\begin{align}
\label{6DGammasDef}
\Gamma^M
=
\left(
\begin{array}{cc}
 0 & \bar{\Sigma}^M{}_{AB}  \\
{\Sigma}^{M,AB} & 0 
\end{array}
\right)
\end{align}
obey the Clifford algebra 
\begin{align}
\{ \Gamma^M , \Gamma^N \} = 2 \eta^{MN} \1_{[8 \times 8]} 
\, ,
\end{align}
with the mostly-minus metric tensor $\eta^{MN} = {\rm diag} (1,-1,-1,-1,-1,-1)$. Their off-diagonal $4 \times 4$ elements are the 
generalized Pauli matrices. The calculations performed in this work required the following set of traces
\begin{alignat}{2}
\ft14 {\rm tr}_4 \left[ \Sigma^M \bar\Sigma^N \right] 
&
= \eta^{MN} &&
\, , \\
\ft14 {\rm tr}_4 \left[ \Sigma^M \bar\Sigma^N \Sigma^K \bar\Sigma^L \right] 
&
= \eta^{MN} 
&&
\eta^{KL} + \eta^{ML} \eta^{NK} - \eta^{MK} \eta^{NL}
\, , \\
\label{6DtraceLC}
\ft14 \tr_4 [\bar\Sigma^{I} \Sigma^{J} \bar\Sigma^{K} \Sigma^{L} \bar\Sigma^{M} \Sigma^{N}]
&
=
\eta^{IJ} 
&&\!\!\!\!
\left[
\eta^{KL} \eta^{MN}
+
\eta^{KN} \eta^{LM}
-
\eta^{KM} \eta^{LN}
\right]
\nonumber\\
&
-
\eta^{IK} 
&&\!\!\!\!
\left[
\eta^{JL} \eta^{MN}
+
\eta^{JN} \eta^{LM}
-
\eta^{JM} \eta^{LN}
\right]
\nonumber\\
&
+
\eta^{IL} 
&&\!\!\!\!
\left[
\eta^{JK} \eta^{MN}
+
\eta^{JN} \eta^{KM} 
-
\eta^{JM} \eta^{KN}
\right]
\nonumber\\
&
-
\eta^{IM} 
&&\!\!\!\!
\left[
\eta^{JN} \eta^{KL}
+
\eta^{JK} \eta^{LN}
-
\eta^{JL} \eta^{KN}
\right]
\nonumber\\
&
+
\eta^{IN} 
&&\!\!\!\!
\left[
\eta^{JM} \eta^{KL}
+
\eta^{JK} \eta^{LM}
-
\eta^{JL} \eta^{KM}
\right]
+
\varepsilon^{IJKLMN}
\, ,
\end{alignat}
where, in the last equation, the Levi-Civita symbol is normalized as $\varepsilon^{012345} = 1$. It is important to realize that the 
six-dimensional chirality has nothing to do with the four-dimensional one!

\section{Four-dimensional MHV unitarity cuts}
\label{Appendix4Dcuts}

For reference purposes, we present in this appendix expressions for the unitarity cuts in Fig.\ \ref{TwoLoopPic} but in four 
dimensions. For the case at hand, all of them can be deduced solely in terms of MHV gluon amplitudes. We write them 
for the helicity assignment ${--++++}$ of $\mathcal{A}_{6} (p_1, p_2, p_3, p_4, p_5, p_6)$ which corresponds to the 
singlet channel for all of the cuts in question. They read, after rationalization and ensuing partial reduction of the Dirac 
strings using on-shellness conditions,
\begin{align}
&
\big. \mathcal{A}^{(2)}_{6} \big|^{--++++}_{{\rm MHV}, {\rm dcut-A}}
\\
&
=
\mathcal{A}^{(0)}_{5} (p_1, p_2, p_3, k_1, k_2) \big|^{--+++}_{\rm MHV}
\mathcal{A}^{(0)}_{4} (-k_2, -k_1, -k_4, -k_3) \big|^{--++}_{\rm MHV}
\mathcal{A}^{(0)}_{5} (k_3, k_4, p_4, p_5, p_6) \big|^{--+++}_{\rm MHV}
\nonumber\\
&\qquad\qquad\qquad\qquad\qquad\quad
=
\frac{- s_{123} \tr_2 [p_1 p_6 k_3 k_4 p_4 p_3 k_1 k_2]}{s_{3 k_1} s_{1 k_2} s_{k_2 k_3} s_{6 k_3} s_{4 k_4}}
\big. \mathcal{A}^{(0)}_{6} \big|^{--++++}_{{\rm MHV}}
\, , \nonumber\\
&
\big. \mathcal{A}^{(2)}_{6} \big|^{--++++}_{{\rm MHV}, {\rm dcut-B}}
\\
&
=
\mathcal{A}^{(0)}_{4} (p_1, p_2, k_1, k_2) \big|^{--++}_{\rm MHV}
\mathcal{A}^{(0)}_{5} (-k_2, -k_1, -k_4, -k_3, p_6) \big|^{--+++}_{\rm MHV}
\mathcal{A}^{(0)}_{5} (k_3, k_4, p_3, p_4, p_5) \big|^{--+++}_{\rm MHV}
\nonumber\\
&\qquad\qquad\qquad\qquad\qquad\quad
=
\frac{\tr_2 [p_3 p_2 p_1 p_6 k_3 k_4 k_1 k_2 p_6 p_5 k_3 k_4]}{s_{2k_1} s_{6,-k_3} s_{6, -k_2} s_{5k_3} s_{3k_4} s_{k_1 k_4}}
\big. \mathcal{A}^{(0)}_{6} \big|^{--++++}_{{\rm MHV}}
\, , \nonumber\\
&
\big. \mathcal{A}^{(2)}_{6} \big|^{--++++}_{{\rm MHV}, {\rm dcut-C}}
\\
&
=
\mathcal{A}^{(0)}_{4} (p_1, p_2, k_1, k_2) \big|^{--++}_{\rm MHV}
\mathcal{A}^{(0)}_{4} (-k_2, -k_1, -k_4, -k_3) \big|^{--++}_{\rm MHV}
\mathcal{A}^{(0)}_{6} (k_3, k_4, p_3, p_4, p_5, p_6) \big|^{--++++}_{\rm MHV}
\nonumber\\
&\qquad\qquad\qquad\qquad\qquad\quad
=
\frac{s_{12} \tr_2 [p_6 p_1 p_2 p_3 k_3 k_4 k_1]}{s_{2k_1} s_{6,k_3} s_{k_1k_4} s_{3k_4}}
\big. \mathcal{A}^{(0)}_{6} \big|^{--++++}_{{\rm MHV}}
\, , \nonumber\\
&
\big. \mathcal{A}^{(2)}_{6} \big|^{--++++}_{{\rm MHV}, {\rm dcut-D}}
\\
&
=
\mathcal{A}^{(0)}_{4} (p_1, p_2, k_1, k_2) \big|^{--++}_{\rm MHV}
\mathcal{A}^{(0)}_{5} (-k_2, -k_1, p_3, -k_4, -k_3, p_6) \big|^{--++++}_{\rm MHV}
\mathcal{A}^{(0)}_{5} (k_3, k_4, p_4, p_5) \big|^{--++}_{\rm MHV}
\nonumber\\
&\qquad\qquad\qquad\qquad\qquad\quad
=
\frac{\tr_2 [p_3 p_2 p_1 p_6 k_2 k_1 p_3 p_4 p_5 p_6 k_3 k_4]}{s_{2k_1} s_{3, -k_1} s_{3,-k_4} s_{6,-k_3} s_{6, -k_2} s_{5k_3} s_{4k_4}}
\big. \mathcal{A}^{(0)}_{6} \big|^{--++++}_{{\rm MHV}}
\, . \nonumber
\end{align}

The above four iterative cuts are the same as in six dimensions, subject to the replacement of Lorentz invariants
$s \to S$! The first three immediately yield the basis integrals and their accompanying coefficients. The last cut D is a 
bit tricky since it lacks the vertical propagators intrinsic to many contributing integrals in the basis. In particular, ten integrals 
$\mathcal{I}^{(2)}_\alpha$ with labels $\alpha = 1, 3, 6, 7, 8, 9, 10, 11, 12, 13$ contribute to it. Notice, however, that the cuts 
A, B, and C unambiguously fix the coefficients of all integrals except for $\mathcal{I}^{(2)}_7$. Thus, we may merely focus 
on extracting just the latter. We can then use the absence of the vertical propagator to our advantage: the heptacut 
of the sum of contributing integrals has to vanish.

\section{Four-dimensional heptacut}
\label{Appendix4Dheptacut}

We turn to the analysis of the helptacut in four dimensions\footnote{All upper-case letter symbols in that figure are understood as 
four-dimensional, $P_i \to p_i$ etc..}, shown in Fig.\ \ref{HeptacutPic}. We adopt the consideration from Ref.\ \cite{Kosower:2010yk} 
of the six-gluon NMHV amplitude (see also Refs.\ \cite{Cachazo:2008hp,Larsen:2012sx} for the leading-singularity analyses) and the
follow-up work on maximal unitarity \cite{Kosower:2011ty}.

The solution for the left and right loop momenta of the hexacut, i.e., cutting internal lines but $q_2$, is parametrized with two complex 
parameters \cite{Kosower:2010yk}, following the spinor decomposition from Ref.\ \cite{delAguila:2004nf,Ossola:2006us},
\begin{align}
\label{4Dq13solutions}
q_1 = z_L \ket{1} [2|
\, , \qquad
q_3 = z_R \ket{4} [5|
\, .
\end{align}
Of course, one can equivalently use the parametrization
\begin{align}
\label{4Dq13solutionsOther}
q_1 = z_L \ket{2} [1|
\, , \qquad
q_3 = z_R \ket{5} [4|
\, ,
\end{align}
or any of the other two alternating picks for $q_{1,3}$ from these two sets. This provides four distinct solutions to the cut conditions for the left and 
right loops, in agreement with \cite{Kosower:2011ty,Larsen:2012sx}. Taking the $q_2$-propagator to its mass shell relates the two variables
$z_L$ and $z_R$ via
\begin{align}
\label{zLzRrelation}
s_{234} + z_L \bra{1} 3 + 4 |2] + z_R \bra{4} 2 + 3 |5] - z_L z_R \vev{14} [25] = 0
\, ,
\end{align}
for the \re{4Dq13solutions}-parametrization, and analogously for the other three. Thus, we can then eliminate $z_L/z_R$ from the right/left 
loops and encode the heptacut in terms of the remaining independent variable.

The heptacut projects on a subset of just five Feynman integrals in Fig.\ \ref{HeptacutPic} with indices $\alpha = 7,10,11,12,13$:
in addition to the sought-after ${\mathcal I}_7$, there are pentaboxes and double-pentagons, which possess additional uncut propagators 
and irreducible numerators. If we choose $z_L$ as the independent degree of freedom, we find for these surviving denominators
\begin{align}
\label{heptacutLSDenominatorsMHV6zL}
s_{3, -k_1} 
&
= s_{23}  - z_L \vev{13} [23]
\, , \\
s_{3, -k_4} 
&
= \vev{23} [34] \frac{\bra{2} 3+4 |5] + z_L \bra{1} 3+4 |5]}{\bra{4} 2 + 3 |5] - z_L \vev{14} [25]}
\, , \nonumber\\
s_{6, -k_2} 
&
= s_{61} + z_L \vev{16} [26]
\, , \nonumber\\
s_{6, -k_3} 
&
= - \vev{16} [56] \frac{\bra{4} 5+6 |1] - z_L \bra{4} 5+6 |2]}{\bra{4} 2 + 3 |5] - z_L \vev{14} [25]}
\, , \nonumber
\end{align}
and the loop momentum-dependent numerators
\begin{align}
\label{heptacutLSNumeratorsMHV6zL}
s_{23, -k_4} 
&
= - z_L  \frac{\vev{23} [34]  \vev{16} [56]}{\bra{4} 2 + 3 |5] - z_L \vev{14} [25]}
\, , \\
s_{34, -k_1} 
&
= s_{234} + z_L \bra{4} 5+6 |1]
\, , \nonumber
\end{align}
on the cut. The numerators in these Eq.\ \re{heptacutLSDenominatorsMHV6zL} reveal the leading singularities of the pentabox and 
double-pentagon Feynman integrals from Fig.\ \ref{TwoLoopIntergalsPic}, in agreement with \cite{Kosower:2010yk}. Equivalently, we 
could have traded $z_L$ in favor of $z_R$ such that
\begin{align}
\label{heptacutLSDenominatorsMHV6zR}
s_{3, -k_1} 
&
= - \vev{34} [23] \frac{\bra{1} 2+3 |4] - z_R \bra{1} 2+3 |5]}{\bra{1} 3 + 4 |2] - z_R \vev{14} [25]}
\, , \\
s_{3, -k_4} 
&
= s_{34}  - z_R \vev{34} [35]
\, , \nonumber\\
s_{6, -k_2} 
&
= 
\vev{16} [56] \frac{\bra{5} 6+1 |2] + z_R \bra{4} 6+1 |2]}{\bra{1} 3 + 4 |2] - z_R \vev{14} [25]}
\, , \nonumber\\
s_{6, -k_3} 
&
= s_{56} + z_R \vev{46} [56]
\, , \nonumber
\end{align}
and
\begin{align}
\label{heptacutLSNumeratorsMHV6zR}
s_{23, -k_4} 
&
= 
s_{234} - z_R \bra{4} 6+1 |5]
\, , \\
s_{34, -k_1} 
&
= 
- z_R  \frac{\vev{34} [23]  \vev{16} [56]}{\bra{1} 3 + 4 |2] - z_R \vev{14} [25]}
\, , \nonumber
\end{align}
respectively. 

In principle, we have to sum over all four solutions to the on-shell kinematical constraint. However, in the current context, they yield 
the same equation for the unknown coefficient $c^{(2)}_7$. So, we focus on just one, in the parametrization \re{4Dq13solutions}. It reads
\begin{align}
\label{HeptaCut4D}
\oint_C \frac{d z_L}{z_L z_R} {\rm Integrand} = 0
\, ,
\end{align}
where $1/(z_L z_R)$ is a Jacobian \cite{Kosower:2011ty,Larsen:2012sx} acquired from integrating out all on-shell propagators by 
global residues \cite{Arkani-Hamed:2009ljj}. It is conventionally referred to as the `composite' leading singularity 
\cite{Buchbinder:2005wp,Cachazo:2008vp}. The vanishing right-hand side in Eq.\ \re{HeptaCut4D} reflects the absence of the 
denominator $s_{3,-k_1,-k_4}$ in the six-gluon MHV amplitude. The integrand on the left-hand side is a combination of 
integrands\footnote{Slightly abusing notations, we designate them by the very same letters as corresponding integrals.} of the 
basis integrals, namely,
\begin{align}
{\rm Integrand}
&
= c_7^{(2)} \mathcal{I}^{(2)}_7
+ 
\sum_{\sigma\cup\sigma^\prime}
\big[
c_{10}^{(2)} \mathcal{I}^{(2)}_{10} + c_{11}^{(2)} \mathcal{I}^{(2)}_{11}
\big]
+
\sum_{\sigma}
\big[
c_{12}^{(2)} \mathcal{I}^{(2)}_{12} + c_{13}^{(2)} \mathcal{I}^{(2)}_{13}
\big]
\, , 
\end{align}
summed over the horizontal and vertical flips, $\sigma = (123456k_1k_2k_3k_4),(543216k_4k_3k_1k_2)$ and 
$\sigma^\prime = (216543k_2k_1k_4k_3),(456123k_3k_4k_2k_1)$. Here, it is tacitly implied that all cut
propagators are stripped down from the integrands, and the leftover loop-momentum numerators/denominators are
substituted according to Eqs.\ (\ref{heptacutLSNumeratorsMHV6zL}--\ref{heptacutLSNumeratorsMHV6zL}) or 
(\ref{heptacutLSDenominatorsMHV6zR}--\ref{heptacutLSNumeratorsMHV6zR}). The heptacut freezes completely
the integrand of $\mathcal{I}^{(2)}_7$ (up the above Jacobian) so that we can replace $\mathcal{I}^{(2)}_7 \to 1$ in 
the first term in Eq.\ \re{HeptaCut4D} and get access to the sought-after coefficient $c_7^{(2)}$. The other contributing 
integrals have unfixed integrands from one or two remaining numerators/propagators. 

There are two possible choices for the integration contour $C$ that pick up the contribution from $\mathcal{I}^{(2)}_7$ 
\cite{Kosower:2011ty}, encircling either $z_L = 0$ or $z_R = 0$ `composite' singularities. Of course, one has to solve 
Eq.\ \re{zLzRrelation} for $z_R$ in terms of $z_L$, for the first one, or the other way around, for the second. Both yield 
identical equations for the expansion coefficient $c^{(2)}_7$. Thus, it suffices to consider just one. Looking at the numerators 
\re{heptacutLSNumeratorsMHV6zL}/\re{heptacutLSNumeratorsMHV6zR}, we observe that $z_L/z_R = 0$ further shrinks the 
list of contributing integrals and their permutations.

For $z_L = 0$, the following equations become valid $k_1 = - p_2$ and $k_2 = - p_1$ since there is no momentum
exchange in the $t$-channel in the left-most amplitude, and thus
\begin{align}
s_{3,-k_1} = s_{23}
\, , \qquad
s_{6,-k_2} = s_{61}
\, , \qquad
s_{34,- k_1} = s_{234}
\, , \qquad
s_{23, -k_4} 
=
0
\, .
\end{align}
Although $s_{3,-k_4}$ and $s_{6, -k_3}$ are much more cumbursome,
\begin{align}
s_{3,-k_4} = s_{34} \frac{\tr_2 [32(6+1)5]}{\tr_2 [34 (6+1) 5]}
\, , \qquad
s_{6,-k_3} = - s_{56} \frac{\tr_2 [4(2+3)16]}{\tr_2 [4(2+3)56]}
\, ,
\end{align}
all terms involving them disappear from the amplitude, thanks to (i) the vanishing numerator $s_{23, -k_4} = 0$, (ii) as well 
as cancellations enabled by the already-fixed integral coefficients $c^{(2)}_{\alpha}$ with $\alpha = 11, 12, 13$. With these 
results at hand, we find that $c^{(2)}_7$ is determined by merely two permutations of the $c^{(2)}_{10}$ and $c^{(2)}_{11}$
coefficients 
\begin{align}
c^{(2)}_7 = - \sum_{
\genfrac{.}{.}{0pt}{2}{(543216)}{(456123)}
} \left[ c^{(2)}_{10} + c^{(2)}_{11} \right]
\, .
\end{align}
The same result is obtained by taking the residue at $z_R = 0$. 

We will adopt the same strategy in the body of the paper for the six-dimensional setup.


\begin{thebibliography}{100}


\bibitem{Bern:1994zx}
Z.~Bern, L.~J. Dixon, D.~C. Dunbar and D.~A. Kosower, \emph{{One loop n point
  gauge theory amplitudes, unitarity and collinear limits}},
  \href{http://dx.doi.org/10.1016/0550-3213(94)90179-1}{\emph{Nucl. Phys. B}
  {\bf 425} (1994) 217--260}, [\href{https://arxiv.org/abs/hep-ph/9403226}{{\tt
  hep-ph/9403226}}].

\bibitem{Bern:1994cg}
Z.~Bern, L.~J. Dixon, D.~C. Dunbar and D.~A. Kosower, \emph{{Fusing gauge
  theory tree amplitudes into loop amplitudes}},
  \href{http://dx.doi.org/10.1016/0550-3213(94)00488-Z}{\emph{Nucl. Phys. B}
  {\bf 435} (1995) 59--101}, [\href{https://arxiv.org/abs/hep-ph/9409265}{{\tt
  hep-ph/9409265}}].

\bibitem{Bern:2004cz}
Z.~Bern, L.~J. Dixon and D.~A. Kosower, \emph{{Two-loop g ---\ensuremath{>} gg
  splitting amplitudes in QCD}},
  \href{http://dx.doi.org/10.1088/1126-6708/2004/08/012}{\emph{JHEP} {\bf 08}
  (2004) 012}, [\href{https://arxiv.org/abs/hep-ph/0404293}{{\tt
  hep-ph/0404293}}].

\bibitem{Berger:2006cz}
C.~F. Berger, Z.~Bern, L.~J. Dixon, D.~Forde and D.~A. Kosower, \emph{{On-shell
  unitarity bootstrap for QCD amplitudes}},
  \href{http://dx.doi.org/10.1016/j.nuclphysbps.2006.09.114}{\emph{Nucl. Phys.
  B Proc. Suppl.} {\bf 160} (2006) 261--270},
  [\href{https://arxiv.org/abs/hep-ph/0610089}{{\tt hep-ph/0610089}}].

\bibitem{Berends:1981rb}
F.~A. Berends, R.~Kleiss, P.~De~Causmaecker, R.~Gastmans and T.~T. Wu,
  \emph{{Single Bremsstrahlung Processes in Gauge Theories}},
  \href{http://dx.doi.org/10.1016/0370-2693(81)90685-7}{\emph{Phys. Lett. B}
  {\bf 103} (1981) 124--128}.

\bibitem{DeCausmaecker:1981wzb}
P.~De~Causmaecker, R.~Gastmans, W.~Troost and T.~T. Wu, \emph{{Helicity
  Amplitudes for Massless QED}},
  \href{http://dx.doi.org/10.1016/0370-2693(81)91025-X}{\emph{Phys. Lett. B}
  {\bf 105} (1981) 215}.

\bibitem{Kleiss:1985yh}
R.~Kleiss and W.~J. Stirling, \emph{{Spinor Techniques for Calculating p anti-p
  ---\ensuremath{>} W+- / Z0 + Jets}},
  \href{http://dx.doi.org/10.1016/0550-3213(85)90285-8}{\emph{Nucl. Phys. B}
  {\bf 262} (1985) 235--262}.

\bibitem{Gunion:1985vca}
J.~F. Gunion and Z.~Kunszt, \emph{{Improved Analytic Techniques for Tree Graph
  Calculations and the G g q anti-q Lepton anti-Lepton Subprocess}},
  \href{http://dx.doi.org/10.1016/0370-2693(85)90774-9}{\emph{Phys. Lett. B}
  {\bf 161} (1985) 333}.

\bibitem{Xu:1986xb}
Z.~Xu, D.-H. Zhang and L.~Chang, \emph{{Helicity Amplitudes for Multiple
  Bremsstrahlung in Massless Nonabelian Gauge Theories}},
  \href{http://dx.doi.org/10.1016/0550-3213(87)90479-2}{\emph{Nucl. Phys. B}
  {\bf 291} (1987) 392--428}.

\bibitem{Cheung:2009dc}
C.~Cheung and D.~O'Connell, \emph{{Amplitudes and Spinor-Helicity in Six
  Dimensions}},
  \href{http://dx.doi.org/10.1088/1126-6708/2009/07/075}{\emph{JHEP} {\bf 07}
  (2009) 075}, [\href{https://arxiv.org/abs/0902.0981}{{\tt 0902.0981}}].

\bibitem{Dennen:2009vk}
T.~Dennen, Y.-t. Huang and W.~Siegel, \emph{{Supertwistor space for 6D maximal
  super Yang-Mills}},
  \href{http://dx.doi.org/10.1007/JHEP04(2010)127}{\emph{JHEP} {\bf 04} (2010)
  127}, [\href{https://arxiv.org/abs/0910.2688}{{\tt 0910.2688}}].

\bibitem{Giele:2008ve}
W.~T. Giele, Z.~Kunszt and K.~Melnikov, \emph{{Full one-loop amplitudes from
  tree amplitudes}},
  \href{http://dx.doi.org/10.1088/1126-6708/2008/04/049}{\emph{JHEP} {\bf 04}
  (2008) 049}, [\href{https://arxiv.org/abs/0801.2237}{{\tt 0801.2237}}].

\bibitem{Bern:2010qa}
Z.~Bern, J.~J. Carrasco, T.~Dennen, Y.-t. Huang and H.~Ita, \emph{{Generalized
  Unitarity and Six-Dimensional Helicity}},
  \href{http://dx.doi.org/10.1103/PhysRevD.83.085022}{\emph{Phys. Rev. D} {\bf
  83} (2011) 085022}, [\href{https://arxiv.org/abs/1010.0494}{{\tt
  1010.0494}}].

\bibitem{Bern:2002tk}
Z.~Bern, A.~De~Freitas and L.~J. Dixon, \emph{{Two loop helicity amplitudes for
  gluon-gluon scattering in QCD and supersymmetric Yang-Mills theory}},
  \href{http://dx.doi.org/10.1088/1126-6708/2002/03/018}{\emph{JHEP} {\bf 03}
  (2002) 018}, [\href{https://arxiv.org/abs/hep-ph/0201161}{{\tt
  hep-ph/0201161}}].

\bibitem{Selivanov:1999ie}
K.~G. Selivanov, \emph{{An Infinite set of tree amplitudes in
  Higgs-Yang-Mills}},
  \href{http://dx.doi.org/10.1016/S0370-2693(99)00760-1}{\emph{Phys. Lett. B}
  {\bf 460} (1999) 116--118}, [\href{https://arxiv.org/abs/hep-th/9906001}{{\tt
  hep-th/9906001}}].

\bibitem{Boels:2010mj}
R.~H. Boels, \emph{{No triangles on the moduli space of maximally
  supersymmetric gauge theory}},
  \href{http://dx.doi.org/10.1007/JHEP05(2010)046}{\emph{JHEP} {\bf 05} (2010)
  046}, [\href{https://arxiv.org/abs/1003.2989}{{\tt 1003.2989}}].

\bibitem{Craig:2011ws}
N.~Craig, H.~Elvang, M.~Kiermaier and T.~Slatyer, \emph{{Massive amplitudes on
  the Coulomb branch of N=4 SYM}},
  \href{http://dx.doi.org/10.1007/JHEP12(2011)097}{\emph{JHEP} {\bf 12} (2011)
  097}, [\href{https://arxiv.org/abs/1104.2050}{{\tt 1104.2050}}].

\bibitem{Brink:1976bc}
L.~Brink, J.~H. Schwarz and J.~Scherk, \emph{{Supersymmetric Yang-Mills
  Theories}}, \href{http://dx.doi.org/10.1016/0550-3213(77)90328-5}{\emph{Nucl.
  Phys. B} {\bf 121} (1977) 77--92}.

\bibitem{Gliozzi:1976qd}
F.~Gliozzi, J.~Scherk and D.~I. Olive, \emph{{Supersymmetry, Supergravity
  Theories and the Dual Spinor Model}},
  \href{http://dx.doi.org/10.1016/0550-3213(77)90206-1}{\emph{Nucl. Phys. B}
  {\bf 122} (1977) 253--290}.

\bibitem{Alday:2009zm}
L.~F. Alday, J.~M. Henn, J.~Plefka and T.~Schuster, \emph{{Scattering into the
  fifth dimension of N=4 super Yang-Mills}},
  \href{http://dx.doi.org/10.1007/JHEP01(2010)077}{\emph{JHEP} {\bf 01} (2010)
  077}, [\href{https://arxiv.org/abs/0908.0684}{{\tt 0908.0684}}].

\bibitem{Caron-Huot:2021usw}
S.~Caron-Huot and F.~Coronado, \emph{{Ten dimensional symmetry of $ \mathcal{N}
  $ = 4 SYM correlators}},
  \href{http://dx.doi.org/10.1007/JHEP03(2022)151}{\emph{JHEP} {\bf 03} (2022)
  151}, [\href{https://arxiv.org/abs/2106.03892}{{\tt 2106.03892}}].

\bibitem{Bork:2022vat}
L.~V. Bork, N.~B. Muzhichkov and E.~S. Sozinov, \emph{{Infrared properties of
  five-point massive amplitudes in $ \mathcal{N} $ = 4 SYM on the Coulomb
  branch}}, \href{http://dx.doi.org/10.1007/JHEP08(2022)173}{\emph{JHEP} {\bf
  08} (2022) 173}, [\href{https://arxiv.org/abs/2201.08762}{{\tt 2201.08762}}].

\bibitem{Belitsky:2025bgb}
A.~V. Belitsky, L.~V. Bork, R.~N. Lee, A.~I. Onishchenko and V.~A. Smirnov,
  \emph{{Five W-boson amplitude = near-null decagon}},
  \href{https://arxiv.org/abs/2510.16471}{{\tt 2510.16471}}.

\bibitem{Belitsky:2022itf}
A.~V. Belitsky, L.~V. Bork, A.~F. Pikelner and V.~A. Smirnov, \emph{{Exact Off
  Shell Sudakov Form Factor in N=4 Supersymmetric Yang-Mills Theory}},
  \href{http://dx.doi.org/10.1103/PhysRevLett.130.091605}{\emph{Phys. Rev.
  Lett.} {\bf 130} (2023) 091605},
  [\href{https://arxiv.org/abs/2209.09263}{{\tt 2209.09263}}].

\bibitem{Belitsky:2023ssv}
A.~V. Belitsky, L.~V. Bork and V.~A. Smirnov, \emph{{Off-shell form factor in $
  \mathcal{N} $=4 sYM at three loops}},
  \href{http://dx.doi.org/10.1007/JHEP11(2023)111}{\emph{JHEP} {\bf 11} (2023)
  111}, [\href{https://arxiv.org/abs/2306.16859}{{\tt 2306.16859}}].

\bibitem{Belitsky:2024agy}
A.~V. Belitsky, L.~V. Bork, J.~M. Grumski-Flores and V.~A. Smirnov,
  \emph{{Three-leg form factor on Coulomb branch}},
  \href{http://dx.doi.org/10.1007/JHEP11(2024)169}{\emph{JHEP} {\bf 11} (2024)
  169}, [\href{https://arxiv.org/abs/2402.18475}{{\tt 2402.18475}}].

\bibitem{Belitsky:2024dcf}
A.~V. Belitsky and L.~V. Bork, \emph{{Off-shell minimal form factors}},
  \href{http://dx.doi.org/10.1007/JHEP07(2025)231}{\emph{JHEP} {\bf 07} (2025)
  231}, [\href{https://arxiv.org/abs/2411.16941}{{\tt 2411.16941}}].

\bibitem{Belitsky:2019fan}
A.~V. Belitsky and G.~P. Korchemsky, \emph{{Exact null octagon}},
  \href{http://dx.doi.org/10.1007/JHEP05(2020)070}{\emph{JHEP} {\bf 05} (2020)
  070}, [\href{https://arxiv.org/abs/1907.13131}{{\tt 1907.13131}}].

\bibitem{Polyakov:1980ca}
A.~M. Polyakov, \emph{{Gauge Fields as Rings of Glue}},
  \href{http://dx.doi.org/10.1016/0550-3213(80)90507-6}{\emph{Nucl. Phys. B}
  {\bf 164} (1980) 171--188}.

\bibitem{Korchemsky:1987wg}
G.~P. Korchemsky and A.~V. Radyushkin, \emph{{Renormalization of the Wilson
  Loops Beyond the Leading Order}},
  \href{http://dx.doi.org/10.1016/0550-3213(87)90277-X}{\emph{Nucl. Phys. B}
  {\bf 283} (1987) 342--364}.

\bibitem{Bern:2008ap}
Z.~Bern, L.~J. Dixon, D.~A. Kosower, R.~Roiban, M.~Spradlin, C.~Vergu et~al.,
  \emph{{The Two-Loop Six-Gluon MHV Amplitude in Maximally Supersymmetric
  Yang-Mills Theory}},
  \href{http://dx.doi.org/10.1103/PhysRevD.78.045007}{\emph{Phys. Rev. D} {\bf
  78} (2008) 045007}, [\href{https://arxiv.org/abs/0803.1465}{{\tt
  0803.1465}}].

\bibitem{Vergu:2009zm}
C.~Vergu, \emph{{Higher point MHV amplitudes in N=4 Supersymmetric Yang-Mills
  Theory}}, \href{http://dx.doi.org/10.1103/PhysRevD.79.125005}{\emph{Phys.
  Rev. D} {\bf 79} (2009) 125005}, [\href{https://arxiv.org/abs/0903.3526}{{\tt
  0903.3526}}].

\bibitem{Vergu:2009tu}
C.~Vergu, \emph{{The Two-loop MHV amplitudes in N=4 supersymmetric Yang-Mills
  theory}}, \href{http://dx.doi.org/10.1103/PhysRevD.80.125025}{\emph{Phys.
  Rev. D} {\bf 80} (2009) 125025}, [\href{https://arxiv.org/abs/0908.2394}{{\tt
  0908.2394}}].

\bibitem{Cachazo:2008hp}
F.~Cachazo, M.~Spradlin and A.~Volovich, \emph{{Leading Singularities of the
  Two-Loop Six-Particle MHV Amplitude}},
  \href{http://dx.doi.org/10.1103/PhysRevD.78.105022}{\emph{Phys. Rev. D} {\bf
  78} (2008) 105022}, [\href{https://arxiv.org/abs/0805.4832}{{\tt
  0805.4832}}].

\bibitem{Arkani-Hamed:2010zjl}
N.~Arkani-Hamed, J.~L. Bourjaily, F.~Cachazo, S.~Caron-Huot and J.~Trnka,
  \emph{{The All-Loop Integrand For Scattering Amplitudes in Planar N=4 SYM}},
  \href{http://dx.doi.org/10.1007/JHEP01(2011)041}{\emph{JHEP} {\bf 01} (2011)
  041}, [\href{https://arxiv.org/abs/1008.2958}{{\tt 1008.2958}}].

\bibitem{Larsen:2012sx}
K.~J. Larsen, \emph{{Global Poles of the Two-Loop Six-Point N=4 SYM
  integrand}}, \href{http://dx.doi.org/10.1103/PhysRevD.86.085032}{\emph{Phys.
  Rev. D} {\bf 86} (2012) 085032}, [\href{https://arxiv.org/abs/1205.0297}{{\tt
  1205.0297}}].

\bibitem{Alday:2007he}
L.~F. Alday and J.~Maldacena, \emph{{Comments on gluon scattering amplitudes
  via AdS/CFT}},
  \href{http://dx.doi.org/10.1088/1126-6708/2007/11/068}{\emph{JHEP} {\bf 11}
  (2007) 068}, [\href{https://arxiv.org/abs/0710.1060}{{\tt 0710.1060}}].

\bibitem{Drummond:2008aq}
J.~M. Drummond, J.~Henn, G.~P. Korchemsky and E.~Sokatchev, \emph{{Hexagon
  Wilson loop = six-gluon MHV amplitude}},
  \href{http://dx.doi.org/10.1016/j.nuclphysb.2009.02.015}{\emph{Nucl. Phys. B}
  {\bf 815} (2009) 142--173}, [\href{https://arxiv.org/abs/0803.1466}{{\tt
  0803.1466}}].

\bibitem{Goncharov:2010jf}
A.~B. Goncharov, M.~Spradlin, C.~Vergu and A.~Volovich, \emph{{Classical
  Polylogarithms for Amplitudes and Wilson Loops}},
  \href{http://dx.doi.org/10.1103/PhysRevLett.105.151605}{\emph{Phys. Rev.
  Lett.} {\bf 105} (2010) 151605}, [\href{https://arxiv.org/abs/1006.5703}{{\tt
  1006.5703}}].

\bibitem{Alday:2007hr}
L.~F. Alday and J.~M. Maldacena, \emph{{Gluon scattering amplitudes at strong
  coupling}},
  \href{http://dx.doi.org/10.1088/1126-6708/2007/06/064}{\emph{JHEP} {\bf 06}
  (2007) 064}, [\href{https://arxiv.org/abs/0705.0303}{{\tt 0705.0303}}].

\bibitem{Drummond:2006rz}
J.~M. Drummond, J.~Henn, V.~A. Smirnov and E.~Sokatchev, \emph{{Magic
  identities for conformal four-point integrals}},
  \href{http://dx.doi.org/10.1088/1126-6708/2007/01/064}{\emph{JHEP} {\bf 01}
  (2007) 064}, [\href{https://arxiv.org/abs/hep-th/0607160}{{\tt
  hep-th/0607160}}].

\bibitem{Drummond:2007aua}
J.~M. Drummond, G.~P. Korchemsky and E.~Sokatchev, \emph{{Conformal properties
  of four-gluon planar amplitudes and Wilson loops}},
  \href{http://dx.doi.org/10.1016/j.nuclphysb.2007.11.041}{\emph{Nucl. Phys. B}
  {\bf 795} (2008) 385--408}, [\href{https://arxiv.org/abs/0707.0243}{{\tt
  0707.0243}}].

\bibitem{Drummond:2007au}
J.~M. Drummond, J.~Henn, G.~P. Korchemsky and E.~Sokatchev, \emph{{Conformal
  Ward identities for Wilson loops and a test of the duality with gluon
  amplitudes}},
  \href{http://dx.doi.org/10.1016/j.nuclphysb.2009.10.013}{\emph{Nucl. Phys. B}
  {\bf 826} (2010) 337--364}, [\href{https://arxiv.org/abs/0712.1223}{{\tt
  0712.1223}}].

\bibitem{Dennen:2010dh}
T.~Dennen and Y.-t. Huang, \emph{{Dual Conformal Properties of Six-Dimensional
  Maximal Super Yang-Mills Amplitudes}},
  \href{http://dx.doi.org/10.1007/JHEP01(2011)140}{\emph{JHEP} {\bf 01} (2011)
  140}, [\href{https://arxiv.org/abs/1010.5874}{{\tt 1010.5874}}].

\bibitem{Huang:2011um}
Y.-t. Huang, \emph{{Non-Chiral S-Matrix of N=4 Super Yang-Mills}},
  \href{https://arxiv.org/abs/1104.2021}{{\tt 1104.2021}}.

\bibitem{Plefka:2014fta}
J.~Plefka, T.~Schuster and V.~Verschinin, \emph{{From Six to Four and More:
  Massless and Massive Maximal Super Yang-Mills Amplitudes in 6d and 4d and
  their Hidden Symmetries}},
  \href{http://dx.doi.org/10.1007/JHEP01(2015)098}{\emph{JHEP} {\bf 01} (2015)
  098}, [\href{https://arxiv.org/abs/1405.7248}{{\tt 1405.7248}}].

\bibitem{Bern:2005iz}
Z.~Bern, L.~J. Dixon and V.~A. Smirnov, \emph{{Iteration of planar amplitudes
  in maximally supersymmetric Yang-Mills theory at three loops and beyond}},
  \href{http://dx.doi.org/10.1103/PhysRevD.72.085001}{\emph{Phys. Rev. D} {\bf
  72} (2005) 085001}, [\href{https://arxiv.org/abs/hep-th/0505205}{{\tt
  hep-th/0505205}}].

\bibitem{Drummond:2007cf}
J.~M. Drummond, J.~Henn, G.~P. Korchemsky and E.~Sokatchev, \emph{{On planar
  gluon amplitudes/Wilson loops duality}},
  \href{http://dx.doi.org/10.1016/j.nuclphysb.2007.11.007}{\emph{Nucl. Phys. B}
  {\bf 795} (2008) 52--68}, [\href{https://arxiv.org/abs/0709.2368}{{\tt
  0709.2368}}].

\bibitem{Brandhuber:2007yx}
A.~Brandhuber, P.~Heslop and G.~Travaglini, \emph{{MHV amplitudes in N=4 super
  Yang-Mills and Wilson loops}},
  \href{http://dx.doi.org/10.1016/j.nuclphysb.2007.11.002}{\emph{Nucl. Phys. B}
  {\bf 794} (2008) 231--243}, [\href{https://arxiv.org/abs/0707.1153}{{\tt
  0707.1153}}].

\bibitem{Basso:2013vsa}
B.~Basso, A.~Sever and P.~Vieira, \emph{{Spacetime and Flux Tube S-Matrices at
  Finite Coupling for N=4 Supersymmetric Yang-Mills Theory}},
  \href{http://dx.doi.org/10.1103/PhysRevLett.111.091602}{\emph{Phys. Rev.
  Lett.} {\bf 111} (2013) 091602}, [\href{https://arxiv.org/abs/1303.1396}{{\tt
  1303.1396}}].

\bibitem{Coronado:2018ypq}
F.~Coronado, \emph{{Perturbative four-point functions in planar $ \mathcal{N}=4
  $ SYM from hexagonalization}},
  \href{http://dx.doi.org/10.1007/JHEP01(2019)056}{\emph{JHEP} {\bf 01} (2019)
  056}, [\href{https://arxiv.org/abs/1811.00467}{{\tt 1811.00467}}].

\bibitem{Coronado:2018cxj}
F.~Coronado, \emph{{Bootstrapping the Simplest Correlator in Planar $\mathcal N
  = 4$ Supersymmetric Yang-Mills Theory to All Loops}},
  \href{http://dx.doi.org/10.1103/PhysRevLett.124.171601}{\emph{Phys. Rev.
  Lett.} {\bf 124} (2020) 171601},
  [\href{https://arxiv.org/abs/1811.03282}{{\tt 1811.03282}}].

\bibitem{Belitsky:2025pnw}
A.~V. Belitsky and V.~A. Smirnov, \emph{{Collinear bootstrap for N=(1,1) sYM}},
  \href{http://dx.doi.org/10.1103/rt29-3dsm}{\emph{Phys. Rev. D} {\bf 112}
  (2025) 105018}, [\href{https://arxiv.org/abs/2503.21915}{{\tt 2503.21915}}].

\bibitem{Belitsky:2024rwv}
A.~V. Belitsky, \emph{{Collinear anatomy}},
  \href{http://dx.doi.org/10.1007/JHEP05(2025)117}{\emph{JHEP} {\bf 05} (2025)
  117}, [\href{https://arxiv.org/abs/2412.11886}{{\tt 2412.11886}}].

\bibitem{Bern:2007ct}
Z.~Bern, J.~J.~M. Carrasco, H.~Johansson and D.~A. Kosower, \emph{{Maximally
  supersymmetric planar Yang-Mills amplitudes at five loops}},
  \href{http://dx.doi.org/10.1103/PhysRevD.76.125020}{\emph{Phys. Rev. D} {\bf
  76} (2007) 125020}, [\href{https://arxiv.org/abs/0705.1864}{{\tt
  0705.1864}}].

\bibitem{Mertig:1990an}
R.~Mertig, M.~Bohm and A.~Denner, \emph{{FEYN CALC: Computer algebraic
  calculation of Feynman amplitudes}},
  \href{http://dx.doi.org/10.1016/0010-4655(91)90130-D}{\emph{Comput. Phys.
  Commun.} {\bf 64} (1991) 345--359}.

\bibitem{Shtabovenko:2023idz}
V.~Shtabovenko, R.~Mertig and F.~Orellana, \emph{{FeynCalc 10: Do multiloop
  integrals dream of computer codes?}},
  \href{http://dx.doi.org/10.1016/j.cpc.2024.109357}{\emph{Comput. Phys.
  Commun.} {\bf 306} (2025) 109357},
  [\href{https://arxiv.org/abs/2312.14089}{{\tt 2312.14089}}].

\bibitem{Kosower:2010yk}
D.~A. Kosower, R.~Roiban and C.~Vergu, \emph{{The Six-Point NMHV amplitude in
  Maximally Supersymmetric Yang-Mills Theory}},
  \href{http://dx.doi.org/10.1103/PhysRevD.83.065018}{\emph{Phys. Rev. D} {\bf
  83} (2011) 065018}, [\href{https://arxiv.org/abs/1009.1376}{{\tt
  1009.1376}}].

\bibitem{Bern:1996ja}
Z.~Bern, L.~J. Dixon, D.~C. Dunbar and D.~A. Kosower, \emph{{One loop selfdual
  and N=4 superYang-Mills}},
  \href{http://dx.doi.org/10.1016/S0370-2693(96)01676-0}{\emph{Phys. Lett. B}
  {\bf 394} (1997) 105--115}, [\href{https://arxiv.org/abs/hep-th/9611127}{{\tt
  hep-th/9611127}}].

\bibitem{Kosower:2011ty}
D.~A. Kosower and K.~J. Larsen, \emph{{Maximal Unitarity at Two Loops}},
  \href{http://dx.doi.org/10.1103/PhysRevD.85.045017}{\emph{Phys. Rev. D} {\bf
  85} (2012) 045017}, [\href{https://arxiv.org/abs/1108.1180}{{\tt
  1108.1180}}].

\bibitem{Mastrolia:2011pr}
P.~Mastrolia and G.~Ossola, \emph{{On the Integrand-Reduction Method for
  Two-Loop Scattering Amplitudes}},
  \href{http://dx.doi.org/10.1007/JHEP11(2011)014}{\emph{JHEP} {\bf 11} (2011)
  014}, [\href{https://arxiv.org/abs/1107.6041}{{\tt 1107.6041}}].

\bibitem{Badger:2012dp}
S.~Badger, H.~Frellesvig and Y.~Zhang, \emph{{Hepta-Cuts of Two-Loop Scattering
  Amplitudes}}, \href{http://dx.doi.org/10.1007/JHEP04(2012)055}{\emph{JHEP}
  {\bf 04} (2012) 055}, [\href{https://arxiv.org/abs/1202.2019}{{\tt
  1202.2019}}].

\bibitem{Buchbinder:2005wp}
E.~I. Buchbinder and F.~Cachazo, \emph{{Two-loop amplitudes of gluons and
  octa-cuts in N=4 super Yang-Mills}},
  \href{http://dx.doi.org/10.1088/1126-6708/2005/11/036}{\emph{JHEP} {\bf 11}
  (2005) 036}, [\href{https://arxiv.org/abs/hep-th/0506126}{{\tt
  hep-th/0506126}}].

\bibitem{Cachazo:2008vp}
F.~Cachazo, \emph{{Sharpening The Leading Singularity}},
  \href{https://arxiv.org/abs/0803.1988}{{\tt 0803.1988}}.

\bibitem{Kotikov:1990kg}
A.~V. Kotikov, \emph{{Differential equations method: New technique for massive
  Feynman diagrams calculation}},
  \href{http://dx.doi.org/10.1016/0370-2693(91)90413-K}{\emph{Phys. Lett. B}
  {\bf 254} (1991) 158--164}.

\bibitem{Gehrmann:1999as}
T.~Gehrmann and E.~Remiddi, \emph{{Differential equations for two-loop
  four-point functions}},
  \href{http://dx.doi.org/10.1016/S0550-3213(00)00223-6}{\emph{Nucl. Phys. B}
  {\bf 580} (2000) 485--518}, [\href{https://arxiv.org/abs/hep-ph/9912329}{{\tt
  hep-ph/9912329}}].

\bibitem{Henn:2013pwa}
J.~M. Henn, \emph{{Multiloop integrals in dimensional regularization made
  simple}}, \href{http://dx.doi.org/10.1103/PhysRevLett.110.251601}{\emph{Phys.
  Rev. Lett.} {\bf 110} (2013) 251601},
  [\href{https://arxiv.org/abs/1304.1806}{{\tt 1304.1806}}].

\bibitem{Belitsky:2024jhe}
A.~V. Belitsky, A.~A. Kokosinskaya, A.~V. Smirnov, V.~V. Voevodin and M.~Zeng,
  \emph{{Efficient Reduction of Feynman Integrals on Supercomputers}},
  \href{http://dx.doi.org/10.1134/S1995080224603709}{\emph{Lobachevskii J.
  Math.} {\bf 45} (2024) 2984--2994},
  [\href{https://arxiv.org/abs/2402.07499}{{\tt 2402.07499}}].

\bibitem{Caron-Huot:2014lda}
S.~Caron-Huot and J.~M. Henn, \emph{{Iterative structure of finite loop
  integrals}}, \href{http://dx.doi.org/10.1007/JHEP06(2014)114}{\emph{JHEP}
  {\bf 06} (2014) 114}, [\href{https://arxiv.org/abs/1404.2922}{{\tt
  1404.2922}}].

\bibitem{Belitsky:2025sin}
A.~V. Belitsky and V.~A. Smirnov, \emph{{Tropical regions of near mass-shell
  pentabox}},  \href{https://arxiv.org/abs/2508.14298}{{\tt 2508.14298}}.

\bibitem{Salvatori:2024nva}
G.~Salvatori, \emph{{The Tropical Geometry of Subtraction Schemes}},
  \href{https://arxiv.org/abs/2406.14606}{{\tt 2406.14606}}.

\bibitem{Goncharov:2009lql}
A.~B. Goncharov, \emph{{A simple construction of Grassmannian polylogarithms}},
   \href{https://arxiv.org/abs/0908.2238}{{\tt 0908.2238}}.

\bibitem{Bork:2025ztu}
L.~V. Bork, R.~N. Lee and A.~I. Onishchenko, \emph{{Method of regions for dual
  conformal integrals}},  \href{https://arxiv.org/abs/2509.12056}{{\tt
  2509.12056}}.

\bibitem{Beneke:1997zp}
M.~Beneke and V.~A. Smirnov, \emph{{Asymptotic expansion of Feynman integrals
  near threshold}},
  \href{http://dx.doi.org/10.1016/S0550-3213(98)00138-2}{\emph{Nucl. Phys. B}
  {\bf 522} (1998) 321--344}, [\href{https://arxiv.org/abs/hep-ph/9711391}{{\tt
  hep-ph/9711391}}].

\bibitem{delAguila:2004nf}
F.~del Aguila and R.~Pittau, \emph{{Recursive numerical calculus of one-loop
  tensor integrals}},
  \href{http://dx.doi.org/10.1088/1126-6708/2004/07/017}{\emph{JHEP} {\bf 07}
  (2004) 017}, [\href{https://arxiv.org/abs/hep-ph/0404120}{{\tt
  hep-ph/0404120}}].

\bibitem{Ossola:2006us}
G.~Ossola, C.~G. Papadopoulos and R.~Pittau, \emph{{Reducing full one-loop
  amplitudes to scalar integrals at the integrand level}},
  \href{http://dx.doi.org/10.1016/j.nuclphysb.2006.11.012}{\emph{Nucl. Phys. B}
  {\bf 763} (2007) 147--169}, [\href{https://arxiv.org/abs/hep-ph/0609007}{{\tt
  hep-ph/0609007}}].

\bibitem{Arkani-Hamed:2009ljj}
N.~Arkani-Hamed, F.~Cachazo, C.~Cheung and J.~Kaplan, \emph{{A Duality For The
  S Matrix}}, \href{http://dx.doi.org/10.1007/JHEP03(2010)020}{\emph{JHEP} {\bf
  03} (2010) 020}, [\href{https://arxiv.org/abs/0907.5418}{{\tt 0907.5418}}].

\end{thebibliography}
\end{document}